\documentclass[twocolumn]{aastex631}
\usepackage{amsmath}

\begin{document}

\title{Modified gravity realizations of quintom dark energy after DESI DR2}

\correspondingauthor{Xin Ren, Emmanuel N. Saridakis and Yi-Fu Cai}
\email{rx76@ustc.edu.cn, msaridak@noa.gr, yifucai@ustc.edu.cn}

\author[0009-0004-7583-1703]{Yuhang Yang}
\affiliation{Department of Astronomy, School of Physical Sciences, University of Science and Technology of China, Hefei 230026, China}
\affiliation{CAS Key Laboratory for Research in Galaxies and Cosmology, School of Astronomy and Space Science, \\
University of Science and Technology of China, Hefei 230026, China}

\author[0000-0003-0194-0697]{Qingqing Wang}
\affiliation{Department of Astronomy, School of Physical Sciences, University of Science and Technology of China, Hefei 230026, China}
\affiliation{CAS Key Laboratory for Research in Galaxies and Cosmology, School of Astronomy and Space Science, \\
University of Science and Technology of China, Hefei 230026, China}
\affiliation{Kavli IPMU (WPI), UTIAS, The University of Tokyo, Kashiwa, Chiba 277-8583, Japan}

\author[0000-0002-5450-0209]{Xin Ren} 
\affiliation{Department of Astronomy, School of Physical Sciences, University of Science and Technology of China, Hefei 230026, China}
\affiliation{CAS Key Laboratory for Research in Galaxies and Cosmology, School of Astronomy and Space Science, \\
University of Science and Technology of China, Hefei 230026, China}

\author[0000-0003-1500-0874]{Emmanuel N. Saridakis} 
\affiliation{National Observatory of Athens, Lofos Nymfon 11852, Greece}
\affiliation{CAS Key Laboratory for Research in Galaxies and Cosmology, School of Astronomy and Space Science, \\
University of Science and Technology of China, Hefei 230026, China}
\affiliation{Departamento de Matem\'{a}ticas, Universidad Cat\'{o}lica del Norte, Avda. Angamos 0610, Casilla 1280, Antofagasta, Chile}

\author[0000-0003-0706-8465]{Yi-Fu Cai} 
\affiliation{Department of Astronomy, School of Physical Sciences, University of Science and Technology of China, Hefei 230026, China}
\affiliation{CAS Key Laboratory for Research in Galaxies and Cosmology, School of Astronomy and Space Science, \\
University of Science and Technology of China, Hefei 230026, China}

\begin{abstract}

We investigate the realization of quintom scenario for dynamical dark energy within   modified gravity theories   that can efficiently fit the recent observational datasets. Starting from a general effective field theory formulation of dark energy in metric-affine geometry, we derive the background action in unitary gauge and we demonstrate how both $f(T)$ and $f(Q)$ gravity can naturally realize quintom behavior through appropriate forms and parameter choices. Additionally, using the Gaussian process reconstruction of the latest DESI DR2 BAO data combined with SNe and CMB observations, we extract the reconstructed dark-energy equation-of-state parameter, showing that it exhibits quintom-type evolution, crossing the phantom divide from below. Moreover, through detailed parameter estimations and application of   information criteria, we compare the model with the quadratic one {and the $\Lambda$CDM model}. Our results show that, due to its rich structure, modified gravity stands as one of the main candidates for the realization of the data-favoured dynamical dark energy. 

\end{abstract}

\section{Introduction}
The accelerated expansion of the Universe was discovered in 1998 through measurements of distances from high-redshift Type Ia supernovae (SNe) \cite{SupernovaSearchTeam:1998fmf, SupernovaCosmologyProject:1998vns}, further confirmed by the Cosmic Microwave Background (CMB) and other cosmological observations. The simplest explanation, a constant dark energy $\Lambda$, was then introduced to describe this acceleration, forming the basis of the standard $\Lambda$CDM cosmological scenario.

However, as cosmological observations have improved in precision, growing evidence suggests that the Universe's expansion may not be driven by a static dark-energy component, but rather by a dynamically evolving one. For example, Planck CMB data combined with weak lensing measurements from the Canada-France-Hawaii Telescope (CFHTLenS)~\cite{Heymans:2012gg} indicate a preference for a dynamical dark energy equation-of-state (EoS), deviating from $\Lambda$CDM at the $2\sigma$ level~\cite{Planck:2015fie, Planck:2018vyg, DES:2017myr}. Further support for dynamical dark energy comes from non-parametric Bayesian reconstructions, which report a $3.5\sigma$ deviation from $\Lambda$CDM~\cite{Zhao:2012aw, Zhao:2017cud, Colgain:2021pmf, Pogosian:2021mcs}. More recently, baryon acoustic oscillation (BAO) measurements from the Dark Energy Spectroscopic Instrument (DESI) suggest dynamical dark energy at $2.5$--$3.9\sigma$ confidence when combined with SNe datasets in last year~\cite{DESI:2024mwx,DESI:2024hhd}. Intriguingly, the DESI~2024 data favors a \textit{quintom}-like behavior~\cite{Feng:2004ad}, where the dark energy EoS parameter crosses the cosmological constant boundary $w = -1$ (also dubbed the phantom divide) from below. This dynamical evolution has received considerable attention and investigation~\cite{Cortes:2024lgw, DESI:2024aqx, DESI:2024kob, Yang:2024kdo, Wang:2024dka, Giare:2024gpk, Mukherjee:2024ryz, Jiang:2024xnu, Dinda:2024kjf, Yang:2025kgc, Giare:2024oil, Liu:2024gfy, Escamilla-Rivera:2024sae, Yin:2024hba, Chudaykin:2024gol, Huang:2025som,RoyChoudhury:2024wri}. The latest DESI Data Release~2, combined with supernova constraints, strengthens this preference up to $4.2\sigma$~\cite{DESI:2025zgx, Lodha:2025qbg,Gu:2025xie}, motivated further investigations into dynamical dark energy~\cite{Ormondroyd:2025iaf, Pang:2025lvh, Anchordoqui:2025fgz, Pan:2025qwy, Pan:2025psn, Paliathanasis:2025dcr}. Meanwhile, the Trans-Planckian Censorship Criterium naturally predicts a time-varying dark energy \cite{Brandenberger:2025hof,Bedroya:2019snp,Li:2025cxn}. These compelling hints, uncovering the fundamental physics behind this dynamical dark energy phenomenon, have become an urgent and pivotal challenge in modern cosmology.

To realize dynamical scenarios of dark energy, one must introduce at least one additional scalar degree of freedom beyond the standard $\Lambda$CDM diagram. The simplest approach involves incorporating a minimally coupled scalar field, which can lead to various dark energy models including quintessence~\cite{Ratra:1987rm, Wetterich:1987fm}, phantom~\cite{Caldwell:1999ew}, or K-essence~\cite{Armendariz-Picon:2000nqq, Chiba:1999ka}. However, as reviewed in~\cite{Cai:2009zp}, these basic single-field models cannot exhibit quintom behavior due to the No-Go theorem, which strictly prohibits the EoS parameter $w$ from crossing the phantom divide in such simple frameworks \cite{Vikman:2004dc,Deffayet:2010qz,Chimento:2008ws}.

In order to overcome this limitation while maintaining a single field description, one must consider more general scalar-tensor theories, such as DHOST~\cite{Langlois:2017mxy,Langlois:2018jdg} and Horndeski theory~\cite{Horndeski:1974wa}. Alternatively, modified gravity theories offer another pathway to dynamical dark energy, where additional gravitational terms can effectively act as dark energy (see~\cite{CANTATA:2021asi} for a comprehensive review). Among these modifications, metric-affine gravity (MAG) has attracted significant attention since it relies solely on spacetime geometry. Similarly to general relativity, MAG encodes gravitational effects through the complete geometric properties of spacetime: curvature, torsion, and non-metricity~\cite{Starobinsky:1980te,Hehl:1994ue,Gronwald:1997bx,JimenezCano:2021rlu,Aoki:2023sum,Capozziello:2002rd,Cai:2015emx,Krssak:2018ywd,BeltranJimenez:2017tkd,Heisenberg:2023lru,Wu:2024vcr}.

The effective field theory (EFT) approach provides a powerful and unified framework for describing various dynamical dark energy models~\cite{Gubitosi:2012hu,Bloomfield:2012ff}. In this paradigm, the additional scalar degree of freedom emerges as a Goldstone boson resulting from the spontaneous symmetry breaking of the time translation in an expanding universe~\cite{Cheung:2007st,Piazza:2013coa}. Remarkably, the EFT framework can encompass both single-scalar field theories and modified gravity theories, including $f(R)$ gravity and Horndeski theory within curvature-based EFT~\cite{Bloomfield:2012ff,Tsujikawa:2014mba}, as well as $f(T)$ gravity in torsion-based EFT~\cite{Li:2018ixg,Yan:2019gbw,Ren:2022aeo}. This unifying framework enables systematic comparisons and investigations of different modified gravity theories under a common theoretical structure.

This manuscript is organized as follows. In Section \ref{sec:eft} we first review the basics of MAG and EFT, then we derive the general EFT action in MAG.
In Section \ref{sec:quintom model} we match the EFT action with $f(T)$ and $f(Q)$ gravity, and introduce the quintom realization. Then, in Section \ref{sec:data&method} we use the current data and Gaussian process to extract  constraints on our quintom model. Finally, we summarize our results in Section \ref{conclusion}.

\section{The effective field theory of dark energy in metric-affine gravity} \label{sec:eft}

In this section we will derive the action of the EFT of dark energy   in the most general metric-affine spacetime. 

\subsection{Metric affine gravity}
First, we briefly review the fundamental geometric quantities in the metric affine spacetime. 
In metric affine gravity (MAG), metric and connection are treated on equal footing, necessitating the use of the Palatini formalism to describe 
gravitational interaction \cite{BeltranJimenez:2019esp}. In this formalism, a general affine connection $\Gamma^{\alpha}_{\ \mu\nu}$ can be decomposed
as
\begin{equation}
 \Gamma^{\alpha}_{\ \mu\nu} = \mathring\Gamma^{\alpha}_{\ \mu\nu}  
+L^{\alpha}_{\ \mu\nu}+K^{\alpha}_{\ \mu\nu},
\end{equation}
where $\mathring\Gamma^{\alpha}_{\ \mu\nu}$ is the Levi-Civita connection, $L ^{\alpha}_{\ \mu\nu}$ and $K^{\alpha}_{\ \mu\nu}$ are the disformation tensor and contortion tensor respectively, characterizing the deviation of 
the full affine connection from the Levi-Civita one. In the following, we use the upper ring to represent that the geometric quantity is 
calculated under the Levi-Civita connection. The affine connection $\Gamma^{\alpha}_{\ \mu\nu}$ establishes the 
affine structure, governing how tensors should be transformed, and defining the covariant derivative $\nabla_{\alpha}$.

Utilizing this general affine connection, we define the basic objects beyond Riemann tensor, namely the non-metricity tensor $Q_{\alpha\mu\nu}$ and the torsion tensor $T^{\alpha}_{\ \mu\nu}$ as
\begin{align}
    Q_{\alpha\mu\nu}&=\nabla_{\alpha}g_{\mu\nu} \\
    T^{\alpha}_{\ \mu\nu}&=\Gamma^{\alpha}_{\ \nu\mu}-\Gamma^{\alpha}_{\ \mu\nu}.
\end{align}
Then the Ricci scalar $R$ in MAG can be written in terms of the Ricci scalar corresponding to the Levi-Civita connection as \cite{Bahamonde:2021gfp,CANTATA:2021asi}
\begin{equation}
    R =\mathring R-Q+T+C+B.
    \label{eq:ricci scalar}
\end{equation}
The non-metricity scalar $Q$, torsion scalar $T$, mixing scalar $C$ and boundary term $B$ are given by
\begin{equation}
    \begin{aligned}
     Q=&\frac{1}{4}Q^{\alpha}Q_{\alpha}- 
\frac{1}{2}\tilde{Q}^{\alpha}Q_{\alpha}-\frac{1}{4}Q_{\alpha\mu\nu}Q^{
\alpha\mu\nu}+\frac{1}{2}Q_{\alpha\mu\nu}Q^{\nu\mu\alpha}, \\
    T=&-T^{\tau}T_{\tau}+\frac{1}{4}T_{\rho\mu\tau}T^{\rho\mu\tau}+\frac{1}{2}T_{\rho\mu\tau}T^{\tau\mu\rho}, \\
    C=&\tilde{Q}_{\rho}T^{\rho}-Q_{\rho}T^{\rho} +Q_{\rho\mu\nu}T^{\nu\rho\mu}, \\
    B=&\mathring\nabla_{\rho}\Big (Q^{\rho}-\tilde{Q}^{\rho}+2T^{\rho}\Big ),
    \end{aligned}
\label{eq:mag scalar}
\end{equation}
where $Q_{\alpha}=g^{\mu\nu}Q_{\alpha\mu\nu}$ and $\tilde 
Q_{\alpha}=g^{\mu\nu}Q_{\mu\alpha\nu}$  represent the two independent traces of 
the non-metricity tensor, and $T^{\mu}=T^{\nu\mu}_{\ \ \ \nu}$ is the only trace of torsion tensor.

\subsection{EFT of MAG}

One of the simplest explanations for dynamical dark energy is to introduce an extra scalar degree of freedom to general relativity (GR), however since this is an ad hoc procedure   the physical origin of this scalar field is   not   clear. One possible explanation is that this extra scalar field arises from the Goldstone field related to the spontaneous symmetry breaking of time translations in an expanding universe, while spatial diffeomorphisms are left unbroken.

The EFT of dark energy provides a systematic way for investigating in a unified framework all dark energy models   as well as modifications of gravity \cite{Gubitosi:2012hu, Bloomfield:2012ff,Creminelli:2008wc, Park:2010cw, Gleyzes:2013ooa}. This formalism describes both the evolution of the cosmological background and the resulting perturbations. A major advantage of the EFT approach lies in its ability to separate the analysis of perturbations from that of the background, enabling independent treatment of each component. These characteristics allow the evolutionary behavior of the three principal types of dark energy to be comprehensively described within the EFT framework.

In the EFT approach, the unitary gauge is conventionally adopted. In this gauge, the scalar degree of freedom is entirely incorporated into the metric. More precisely, the temporal coordinate is defined as a function of the scalar field itself, resulting in the vanishing of field fluctuations around the background $\delta \phi(t,\mathbf{x}) \equiv \phi(t,\mathbf{x}) - \bar{\phi}(t) = 0$, where $\bar{\phi}(t)$ denotes the background value of the scalar field. To restore both the scalar degree of freedom and full diffeomorphism invariance, one may employ the ``Stueckelberg'' trick. This is achieved through an infinitesimal time diffeomorphism transformation $t \mapsto t + \pi(x)$, where the field $\pi(x)$ serves as the dynamical perturbation that governs the scalar sector of dark energy. Through this procedure, the dynamical role of the scalar field is naturally reinstated within the EFT framework, yielding a complete description of the physical system.

We proceed in  constructing the  EFT formalism in the most general metric affine spacetime. Working in the unitary gauge, the operators appearing in the general EFT action should be invariant under the residual symetries of unbroken spatial diffeomorphisms. In general, it should contain \cite{Cheung:2007st}:
\begin{enumerate}
    \item Four-dimensional diff-invariant scalars multiplied by functions of time.
    \item Four-dimensional covariant tensors with free upper $0$ indices, where all spatial indices must be contracted.
    \item Three-dimensional terms belonging to the $t=const.$ hypersurface.
\end{enumerate}
Thus, it is   straightforward     to define the unit timelike-vector $n_{\mu}$ normal to the hypersurface slicing by the scalar field $\phi$, namely 
\begin{equation}
    n_{\mu}=-\frac{\partial_{\mu}\phi(t)}{\sqrt{-(\partial \phi)^2}}=-\frac{\delta_{\mu}^{\ 0}}{\sqrt{-g^{00}}}.
\end{equation}
The induced metric $\gamma_{\mu\nu}$ can be defined as $\gamma_{\mu\nu}=g_{\mu\nu}+n_{\mu}n_{\nu}$, while the covariant derivatives of $n_{\mu}$ should   be considered, too. Moreover, we define the gradient tensor in MAG as
\begin{equation}
    \mathcal{K}_{\mu\nu}= \gamma^{\rho}_{\ \mu} \nabla_{\rho}n_{\nu},
\end{equation}
and we  can rewrite  it   using the relation between general affine connection and Levi-Civita connection, resulting to
\begin{equation}
\mathcal{K}_{\mu\nu}=\mathring{K}_{\mu\nu}-G^{\rho}_{\ \mu\nu}n_{\rho}+\frac{1}{2g^{00}}n_{\mu}Q_{\nu}^{\ 00},
\label{eq:gradient tensor}
\end{equation}
where $\mathring{K}_{\mu\nu}=\gamma^{\rho}_{\ \mu} \mathring\nabla_{\rho}n_{\nu}$ is the extrinsic curvature tensor in Levi-Civita connection, and $G^{\rho}_{\ \mu\nu}=L^{\alpha}_{\ \mu\nu}+K^{\alpha}_{\ \mu\nu}$. One can easily check that $\mathcal{K}_{\mu\nu}n^{\mu}=\mathcal{K}_{\mu\nu}n^{\nu}=0$, which implies that $\mathcal{K}_{\mu\nu}$ is orthogonal to $n^{\mu}$, belonging to the three-dimensional hypersurface. Additionally, the trace of the gradient tensor can be written as
\begin{equation}
\begin{aligned}
        \mathcal{K}=\mathring{K} + \frac{1}{2\sqrt{-g^{00}}}\Big (Q^0 \!-\!2 \tilde{Q}^0\!+\!2T^{0} \Big ) +\frac{1}{2(\sqrt{-g^{00}})^3}Q^{000} .
    \label{eq:trace gradient tensor}
\end{aligned}
\end{equation}
From Eq.~\eqref{eq:gradient tensor} and Eq.~\eqref{eq:trace gradient tensor} we can see that the gradient tensor $\mathcal{K}_{\mu \nu}$ or its trace $\mathcal{K}$ with general affine connection, can be written as a combination of $\mathring K_{\mu \nu}$ or $\mathring K$ with other four-dimensional covariant tensors with free upper $0$ indices.

Let us now consider an operator built by the contraction of two tensors $X$ and $Y$. By expanding $X=X^{(0)}+\delta X$ and $Y=Y^{(0)}+\delta Y$, we have
\begin{equation}
    XY=\delta X \delta Y+X^{(0)}Y+XY^{(0)}-X^{(0)}Y^{(0)}.
\end{equation}
The first term is quadratic in perturbation and therefore we keep it. Furthermore, we assume that $X$ or $Y$ is linear in $\mathring R_{\mu\nu\rho\sigma}$, $\mathring{K}_{\mu\nu}$, $T^{\rho}_{\ \mu\nu}$, $Q_{\alpha\mu\nu}$ and $\mathcal{K}_{\mu\nu}$, with covariant derivatives acting on them. Due to the relation between the covariant derivative $\nabla$ and $\mathring \nabla$, we can always decomposed $\nabla$ into $\mathring \nabla$ along with a term contracted with $T^{\alpha}_{\ \mu\nu}$ and $Q^{\alpha}_{\ \mu\nu}$. Hence, it is sufficient to  consider only the covariant derivatives $\mathring \nabla$ with respect to the Levi-Civita connection, which implies that we can use integrations by parts to absorb the $\mathring \nabla$. Additionally, the only possible scalar terms will be $\mathring{K}$, $\mathring{R}$, $\mathring R^{00}$, $\mathcal{K}$, $T^{0}$, $Q^{0}$, $\tilde{Q}^{0}$, $T$, $Q$, $C$ and $B$, and through relation Eq.~\eqref{eq:trace gradient tensor}  we can eliminate $\mathcal{K}$ in terms of $T^{0}$, $Q^{0}$, $\tilde{Q}^0$ and $Q^{000}$. The integration of $\mathring R^{00}$ and $\mathring {K}$ with time-dependent coefficients gives just the linear operator $g^{00}$, along with some invariant terms that start quadratic in perturbations, and hence we can avoid them in the background action by using $g^{00}$ instead. Finally, note that the integration of the boundary term $B$ with a time-dependent coefficient becomes
\begin{equation}
    \begin{aligned}
    \int d^4x\sqrt{-g}m(t)B=&\int d^4x\sqrt{-g}m(t)\mathring\nabla_{\rho}\Big (Q^{\rho}-\tilde{Q}^{\rho}+2T^{\rho}\Big ) \\
    =&\int d^4x\sqrt{-g} \Dot{m}(t)\Big( Q^{0}-\tilde{Q}^{0}+2T^{0} \Big).
    \end{aligned}
\end{equation}
Therefore, we can adsorb the boundary term into $Q^{0}$, $\tilde{Q}^{0}$ and $T^{0}$.
Lastly, for the matter sector we  assume that the weak equivalence principle (WEP) is valid, and thus   the matter field $\psi_{m}$ is minimally coupled to the metric $g_{\mu\nu}$ through the action $S_m[g_{\mu\nu};\psi_{m}]$, i.e. we will work in the Jordan frame.

In summary, we can now write   the most general EFT action in MAG as
\begin{equation}
    \begin{aligned}
    S=&\int d^4x\sqrt{-g}\Bigg[ \frac{M_P^2}{2}\Big (\Psi(t)\mathring R+d(t)T+e(t)Q\Big ) \\
    &+\frac{M_P^2}{2}\Big (g(t)T^0+h(t)Q^0+j(t)\tilde{Q}^0 \Big)\\
    &-\Lambda(t)-b(t)g^{00}-k(t)Q^{000} +\frac{M_P^2}{2}m(t)C \Bigg]+S_{DE}^{(2)},
    \end{aligned}
    \label{eq:EFT MAG}
\end{equation}
with $M_p^2=1/8\pi G$  the Planck mass, and where $\Psi$, $\Lambda$, $d$, $e$, $g$, $h$, $j$, $b$, $k$ and $m$ are functions of the time coordinate $t$. Additionally, 
$S_{DE}^{(2)}$ indicates terms that are explicitly quadratic in   perturbations and therefore do not affect the background.

\section{Quintom dark energy within MAG} \label{sec:quintom model}

The $w_0-w_a$ parameterization was originally proposed as a phenomenological tool to describe dark energy dynamics near the present epoch. The remarkable effectiveness of this simple first-order parameterization in matching the latest observed cosmic expansion history remains theoretically intriguing. In this section we explore possible physical interpretations  of such parameterizations, and moving beyond their purely phenomenological origins  we investigate the quintom realization in the framework of metric affine  gravity. As usual, 
in order to apply MAG at a cosmological  framework we impose the spatially-flat Friedmann-Robertson-Walker 
(FRW) metric of the form $
ds^{2}=-dt^{2}+a^{2}(t)\delta_{ij}dx^{i}dx^{j}$, with $a(t)$ the scale factor.

\subsection{$f(T)$ gravity}

In $f(T)$ gravity \cite{Cai:2015emx}  the geometry is flat and metric-compatible, and thus   the general relation between $R$ and $\mathring R$, Eq.~\eqref{eq:ricci scalar}, can be rewritten as 
\begin{equation}
    \mathring R=-T-2\mathring \nabla_{\rho}T^{\rho}.
    \label{eq:R_T relation}
\end{equation}
In the context of $f(T)$ gravity we adopt a time slicing that aligns with uniform hypersurfaces of the torsion scalar $T$. This gauge choice is particularly advantageous since it ensures that all higher-order terms ($\mathcal{O}(\delta T^n)$ with $n \geq 2$) in the expansion of the $f(T)$ action around the background value $T^{(0)}$, vanish identically. This simplification occurs due to the fact that any such nonlinear contributions to the equations of motion must necessarily contain at least one factor of $\delta T \equiv T - T^{(0)}$, which by construction vanishes in this gauge. Consequently, in the unitary gauge where $\delta T = 0$, the action retains only linear terms in the perturbation variables.
Thus, in the unitary gauge we have
\begin{equation}
    f(T) \stackrel{unitary}{\longrightarrow}  f_T(T^{(0)})T+f(T^{(0)})-f_T(T^{(0)})T^{(0)}.
\end{equation}
Using Eq.~\eqref{eq:R_T relation} and integrating by parts to drop the boundary term, we finally obtain
\begin{equation}
    \begin{aligned}
        f(T) \stackrel{unitary}{\longrightarrow} & -f_T(T^{(0)}) \mathring{R}+2\dot{f_T}(T^{(0)})T^{(0)} \\
    &-T^{(0)}f_T(T^{(0)})+f(T^{(0)}).
    \end{aligned}
\end{equation}
Comparing with Eq.~\eqref{eq:EFT MAG}, we   observe that the non-zero terms are
\begin{equation}
    \begin{aligned}
        \Psi(t)&=-f_T(T^{(0)}), \quad d(t)=2\dot{f_T}(T^{(0)}) \\
        \Lambda(t)&=-\frac{M_p^2}{2}\Big[ f(T^{(0)})-T^{(0)}f_T(T^{(0)}) \Big ],
    \end{aligned}
\end{equation}
which are consistent with the expressions in \cite{Li:2018ixg}.
In $f(T)$ cosmology  we can define  the effective dark energy pressure and energy density as
\begin{equation}
\begin{aligned}
p_{f(T)}=&\frac{f-Tf_T+2T^2f_{TT}}{2f_T+4Tf_{TT}} ~ , \\
 \rho_{f(T)}=&Tf_T-\frac{1}{2}(f+T) ~, 
\end{aligned}
\end{equation}
with $f_T={\rm d}f/{\rm d}T$, $f_{TT}={\rm d}^2f/{\rm d}T^2$ and $T=6H^2$. Therefore, 
the effective EoS of dark energy is given by
\begin{equation}
\label{wofg}
w_{f(T)} \equiv \frac{p_{f(T)}}{\rho_{f(T)}}=\frac{f-Tf_{T}+2 T^2 f_{T T}}{\left[f_{T}+2 T f_{T T}\right]\left[2T f_{T}-f-T\right]}. 
\end{equation}

\subsection{$f(Q)$ gravity}

 In $f(Q)$ gravity  the geometry is flat and torsion-less \cite{Nester:1998mp}. The relation between $\mathring R$ and $Q$ arises from Eq.~\eqref{eq:ricci scalar} as 
\begin{equation}
    \mathring R=Q-\mathring\nabla_{\rho}\Big (Q^{\rho}-\tilde{Q}^{\rho}).
\end{equation}
For simplicity, in $f(Q)$ cosmology around an FRW spacetime   we choose the coincident gauge with the simplest connection that inherits the symmetries of the
background spacetime     \cite{BeltranJimenez:2019tme,Heisenberg:2023lru},  which leads to $Q=-6H^2$. However, in more general cases, there are two different branches beyond the coincident gauge where a free temporal function $\gamma(t)$ appears, and thus the evolution of $Q$ is not monotonous anymore \cite{Hohmann:2019fvf,Yang:2024tkw} (see \cite{Basilakos:2025olm,Paliathanasis:2023pqp} for the quintom realization in such general-connection $f(Q)$ frameworks). The non-monotonicity of $Q$ implies that we can no longer simply choose a constant-$Q$ hypersurface as our time slicing. Since these non-trivial branches are beyond the consideration of this work,  in the following we  focus on the coincident gauge, in which we can safely choose the time slicing to coincide with our constant-$Q$ hypersurface.

Similarly to the steps we followed in  $f(T)$ gravity, we rewrite the $f(Q)$ action in the unitary gauge as
\begin{equation}
    f(Q)\stackrel{unitary}{\longrightarrow} f_Q(Q^{(0)})Q+f(Q^{(0)})-f_Q(Q^{(0)})Q^{(0)}.
\end{equation}
Replacing $Q$ with $\mathring R$ and integrating by parts, we acquire
\begin{equation}
    \begin{aligned}
    f(Q)\stackrel{unitary}{\longrightarrow}&f_Q(Q^{(0)})\mathring R+\Dot{f_Q}(Q^{(0)})(\tilde{Q}^0-Q^0)\\
    &-Q^{(0)}f_Q(Q^{(0)})+f(Q^{(0)}).
    \end{aligned}
\end{equation}
As we see, the non-zero terms are
\begin{equation}
    \begin{aligned}
        &\Psi(t)=f_Q(Q^{(0)}), \quad j(t)=-h(t)=\Dot{f_Q}(Q^{(0)}), \\
        &\Lambda(t)=-\frac{M_p^2}{2}\Big [f(Q^{(0)})-Q^{(0)}f_Q(Q^{(0)}) \Big ].
    \end{aligned}
\end{equation}
Finally, we mention that for $f(Q)$ cosmology in the  coincident gauge, the  background evolution is the same as in $f(T)$ gravity.

\subsection{Quintom Realization}

\begin{figure*}[htbp]
    \centering
    \begin{minipage}{0.48\textwidth}
        \includegraphics[width=\textwidth]{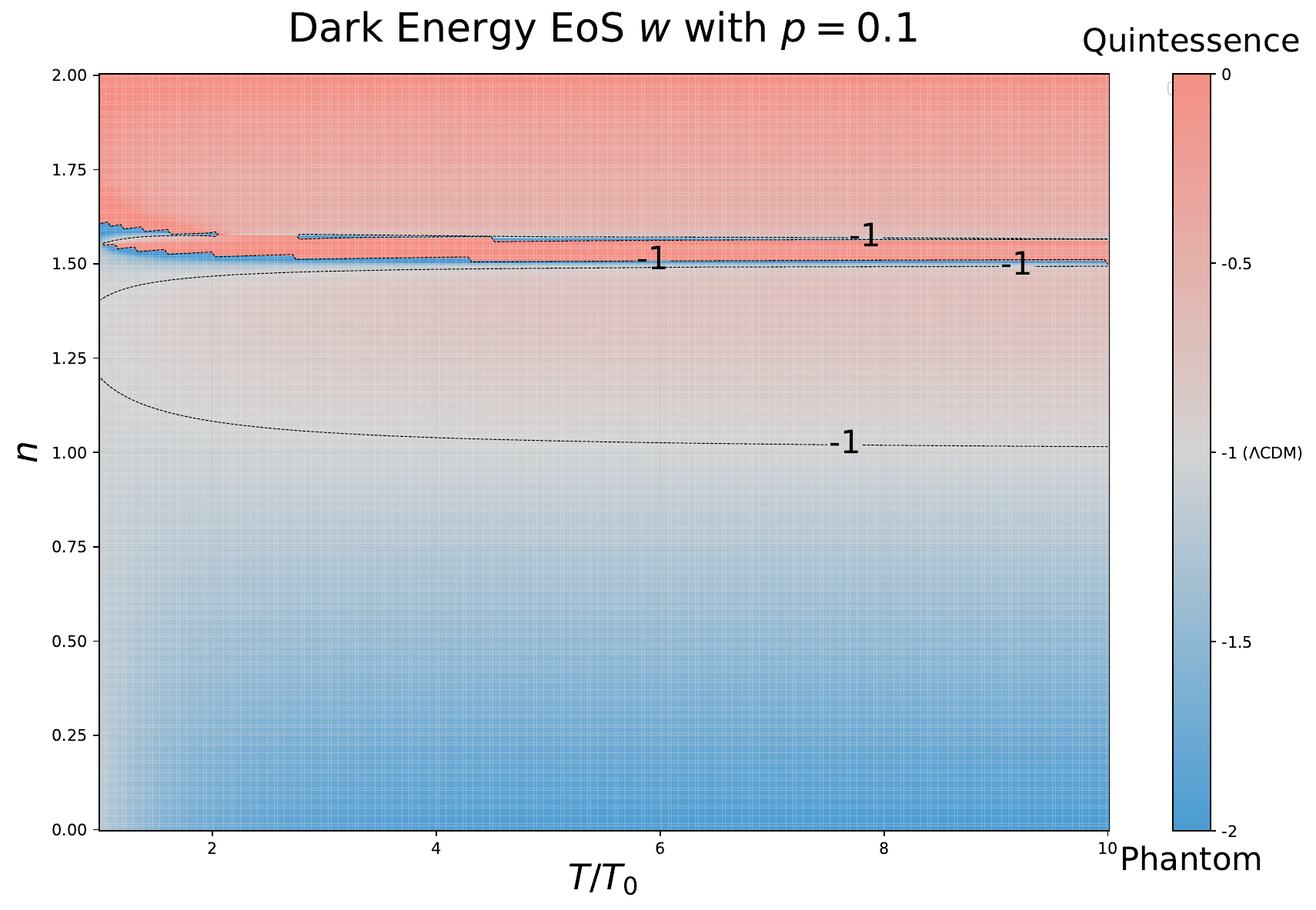}
        \label{fig:w_0.1}
    \end{minipage}
    \hfill
    \begin{minipage}{0.48\textwidth}
        \includegraphics[width=\textwidth]{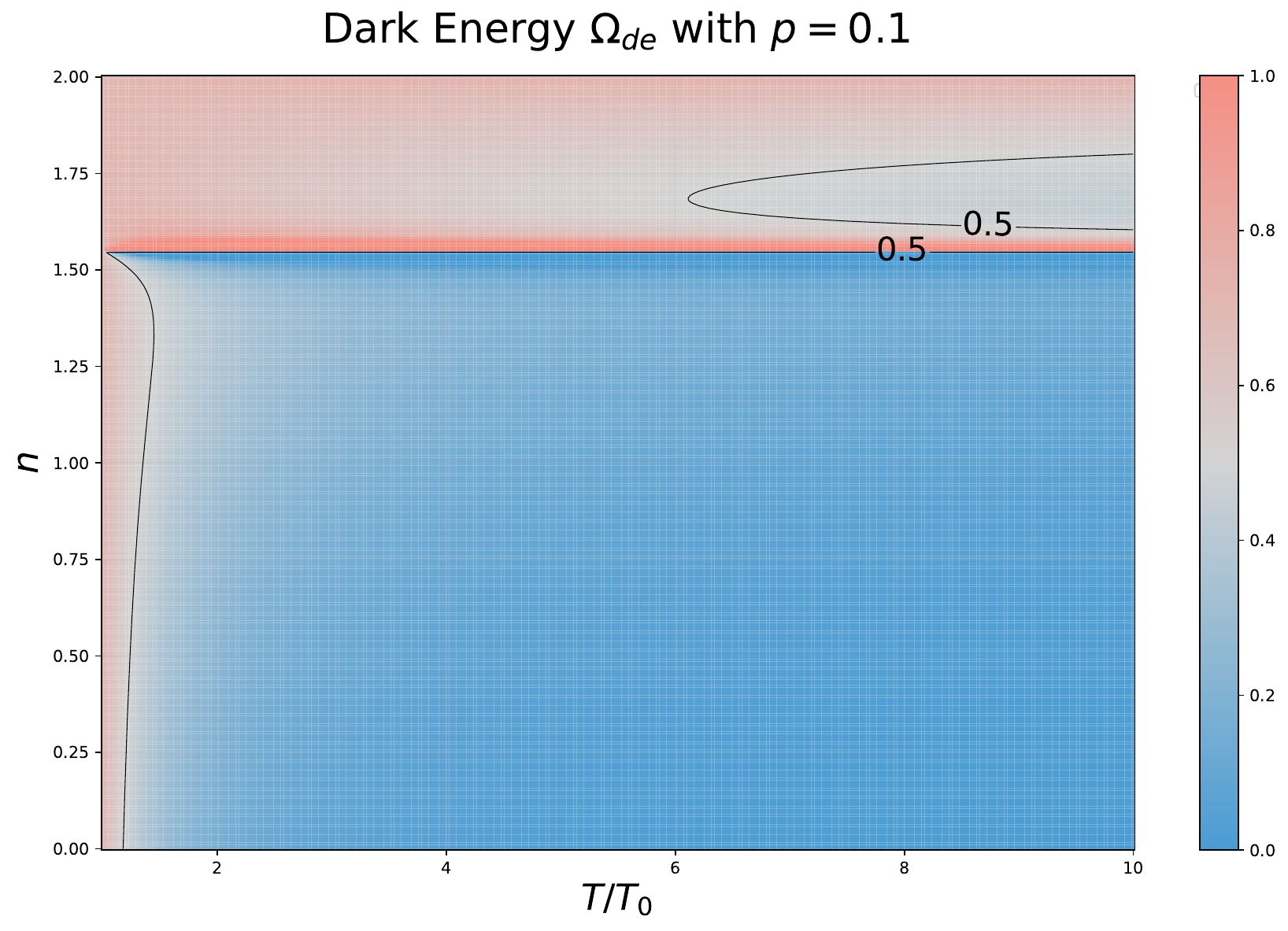}
        \label{fig:omega_0.1}
    \end{minipage}
    
    \vspace{0.2cm}
    
    \begin{minipage}{0.48\textwidth}
        \includegraphics[width=\textwidth]{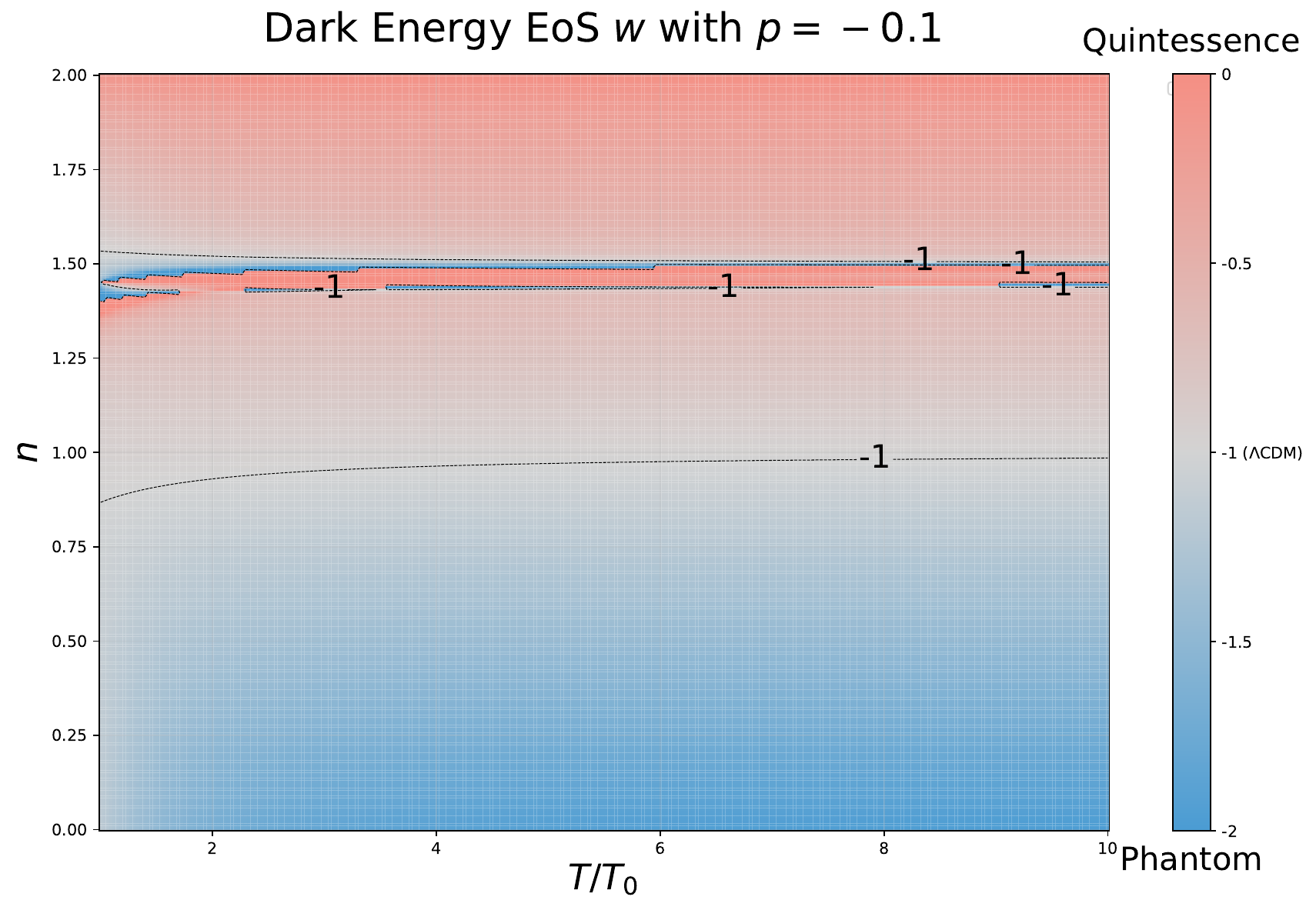}
        \label{fig:w_-0.1}
    \end{minipage}
    \hfill
    \begin{minipage}{0.48\textwidth}
        \includegraphics[width=\textwidth]{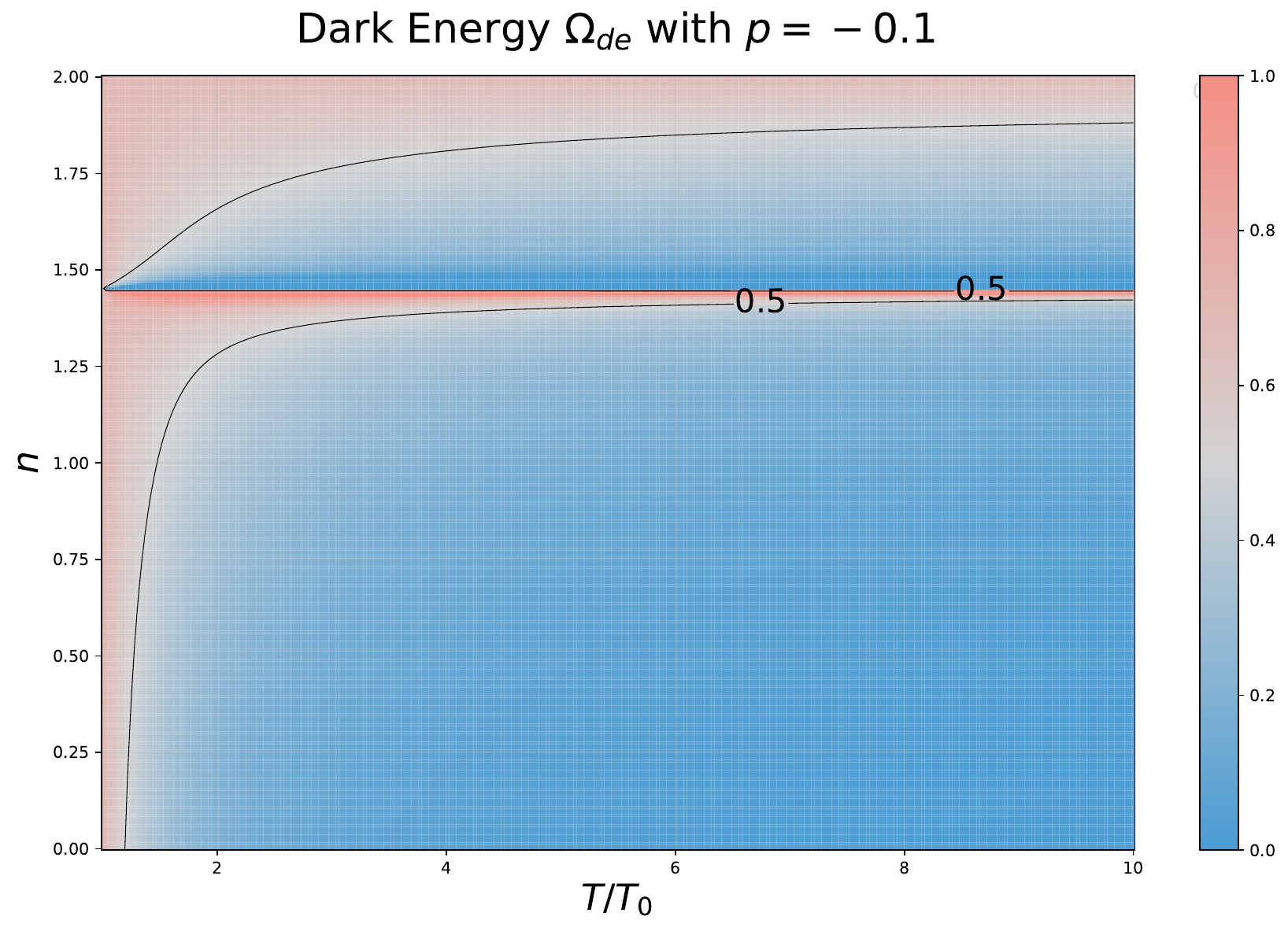}
        \label{fig:omega_-0.1}
    \end{minipage}
    
    \caption{\it{The      dark energy EoS parameter $w$ and the dark energy density parameter $\Omega_{de}$ under different parameters choices. We have imposed $\Lambda=0$, $\Omega_{m,0}=0.3$ and $H_0=70$ $\rm km$ $\rm s^{-1}Mpc^{-1}$.}}
    \label{fig:combined_results}
\end{figure*}

Let us now consider a specific $f(T)$ model that can lead to the realization of the quintom behavior. Since in the  coincident gauge the  background evolution of $f(Q)$ cosmology coincides with that of $f(T)$ cosmology, the following results hold for  $f(Q)$ gravity too, under the substitution $T\rightarrow Q$.

We choose
\begin{equation}
    f(T)=T+\alpha (-T)^n\Big[1-e^{pT_0/T}\Big]-2\Lambda ,
    \label{eq:quintom ft model}
\end{equation}
where $\alpha$, $n$ and $p$ are parameters, and where $T_0=6H_0^2$ with $H_0$     the Hubble parameter at present. We mention that among these three parameters, only two are   independent, while the third one  is eliminated by applying the first Friedmann equation at present time. Without loss of generality  we choose to treat $n$ and $p$ as the free parameters of our analysis, and consequently  the parameter $\alpha$ can be expressed   as
\begin{equation}
    \alpha= \frac{[1-\Omega_{m,0}-\Lambda/(3H_0^2)](6H_0)^{1-n}}{2n-1+(1-2n+2p)e^{p})},
\end{equation}
where $\Omega_{m,0}$ refer to the matter density parameter at present. Finally, note that for $n=1$ or $p=0$, model (\ref{eq:quintom ft model})  reduces respectively to the power-law   and exponential models discussed in \cite{Linder:2010py}. The motivation for their combination in 
(\ref{eq:quintom ft model})  is that   these simple cases alone cannot achieve the phantom-divide crossing and experience the   quintom behavior.
 
The dark-energy equation-of-state parameter  (\ref{wofg}) becomes
 \begin{widetext}\small{}
\begin{equation}
    w(z)=\frac{\frac{-2\Lambda(6H_0)^n}{x^n}+\alpha\left\{-(n-1)(2n-1)+e^{p/x}\Big [(n-1)(2n-1)+(5-4n)\frac{p}{x}+2\frac{p^2}{x^2} \Big ] \right\}}{\left\{\frac{-2\Lambda(6H_0)^n}{x^n}+\alpha \Big [1-2n+e^{p/x}(2n-1-2\frac{p}{x}) \Big ] \right\}\Big  [1+\alpha (6H_0^2)^{n-1}x^{n-1}((1-2n)n+e^{p/x}(n(2n-1)+(3-4n)\frac{p}{x}+2\frac{p^2}{x^2})) \Big ]},
\end{equation}
\end{widetext}
where $x=T(z)/T_0=E^2(z)$. Additionally,    the dark-energy density parameter is given by
\begin{eqnarray}
 && \!\!\!\!\!\!\!\!\!\!\!\!\!\!\!\!\! \!\!\!\!\!  \Omega_{de}(z)=\frac{\Lambda}{3xH_0^2}+\alpha (6H_0^2)^{n-1}x^{n-1} \nonumber \\
   & &\ \ \ \ \ \ \ \ \ \ \ \ \ \cdot \Big [ 2n-1+ e^{p/x}(1-2n+2\frac{p}{x}) \Big ].
 \end{eqnarray}

In order to recover the standard $\Lambda$CDM paradigm, we require that the dark energy should be negligible in the early universe. The evolution of $w(z)$ and $\Omega_{de}(z)$ with respect to   $T/T_0$ is shown in Fig. \ref{fig:combined_results}. If we set $\Lambda=0$, we find that in order to obtain a quintom realization, $p$ and $n-1$ should have the same sign. In particular, for $p<0$  the model can exhibit   quintom-B dynamics, while for $p>0$ it corresponds to  quintom-A evolution. Finally, the condition $ \lim_{z\to \infty}\Omega_{de}(z)$ can   impose a constraint on the parameter space. For instance, for $p<0$, $n$ should satisfy $n<1$ and $w_{T,0}>-1$ to ensure a quintom-B evolution in the past, where  $w_{T,0}$ is the current value of dark energy EoS parameter. Similarly,  for $p>0$, $n>1$ together with $w_{T,0}<-1$, can result in a quintom-A history.

\begin{figure*}[htbp]
    \centering
    \begin{minipage}{0.48\textwidth}
        \includegraphics[width=\textwidth]{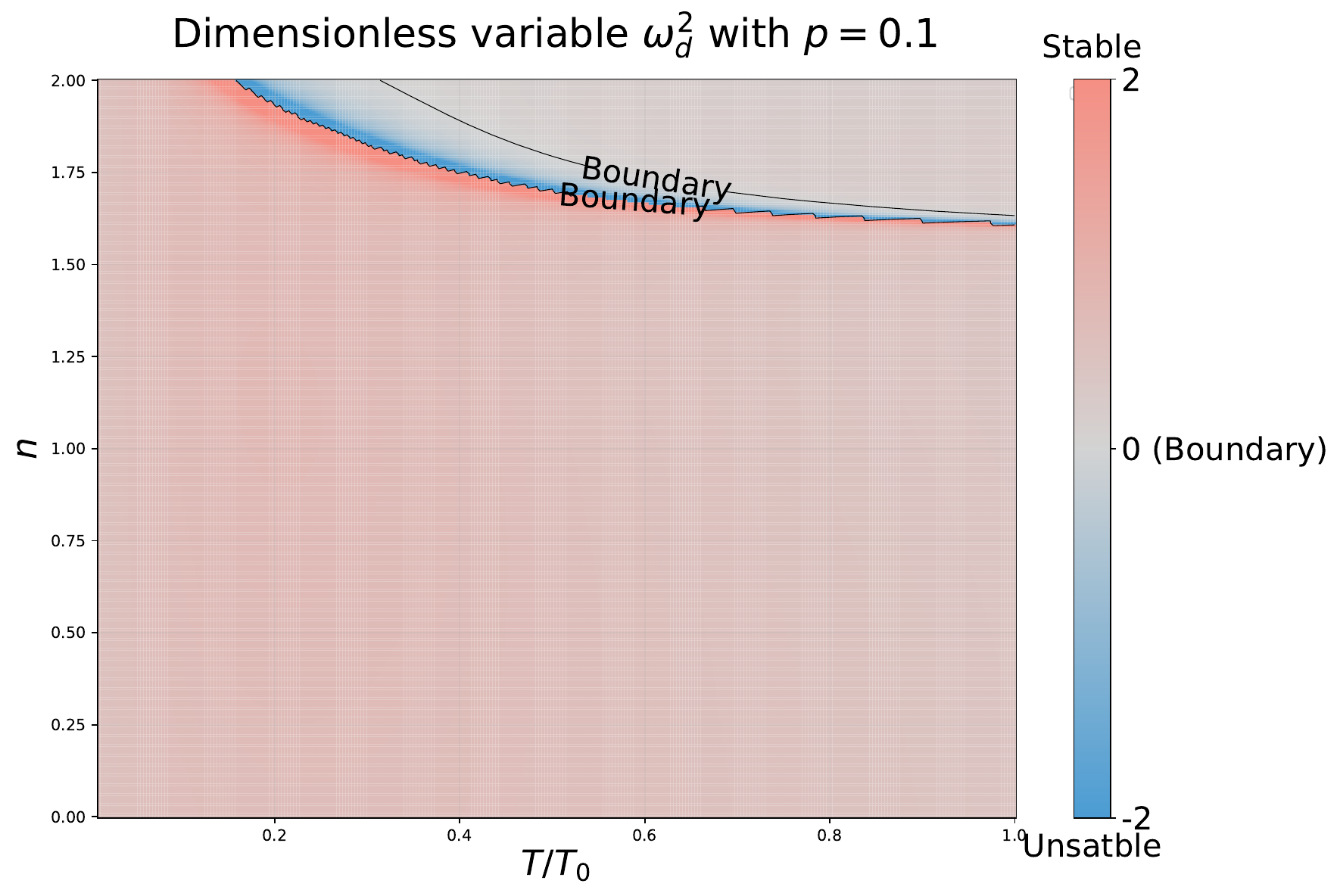}
    \end{minipage}
    \hfill
    \begin{minipage}{0.48\textwidth}
        \includegraphics[width=\textwidth]{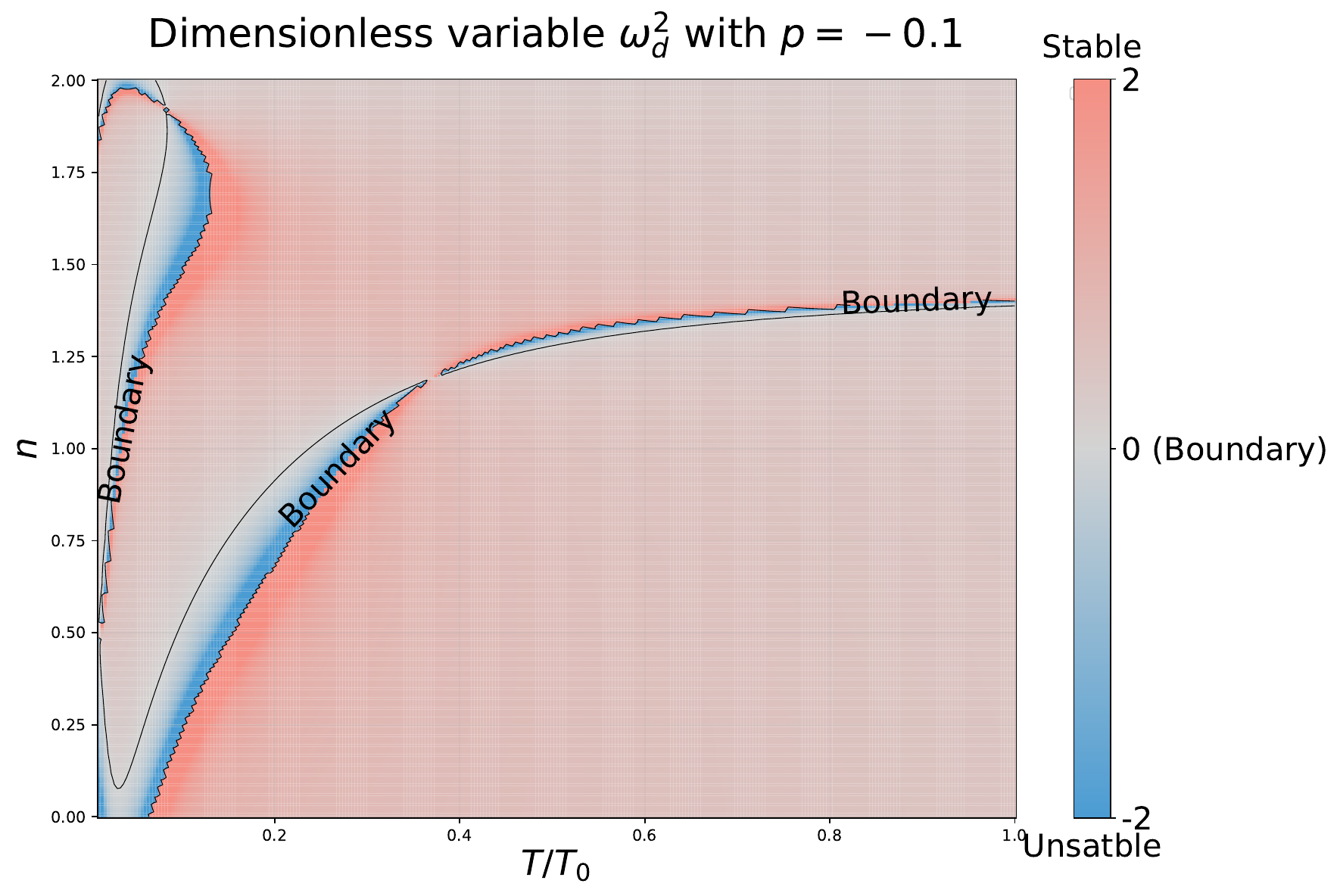}
    \end{minipage}
    \caption{\it{The dimensionless perturbation variable $\omega_d^2$ under different parameters choices. We have imposed $\Lambda=0$, $\Omega_{m,0}=0.3$ and $H_0=70$ $\rm km$ $\rm s^{-1}Mpc^{-1}$.}}
    \label{fig:omega2}
\end{figure*}

However, recalling the No-Go theorem for the scalar field, it is not enough to study only the background evolution of the model, and we will next study the stability of the theory from the aspect of scalar perturbation. We work in the Newtonian gauge, namely
\begin{equation}
    ds^2=(1+2\Psi)dt^2-a^2(1-2\Phi)\delta_{ij}dx^idx^j.
\end{equation}
In the following, we will focus on the perturbation of the pure gravitational sector, ignoring the effect of matter sector and focusing on the dark-energy-domination era. On the one hand, to make it easy to calculate, and on the other hand, if the theory is already unstable in the pure gravitational part, then adding matter does not eliminate the instability. Suppose there are no anisotropic stress, thus we have $\Psi=\Phi$. Working in Fourier space, the evolution equation of $\Phi_k$ is given by \cite{Chen:2010va}
\begin{equation}
    \Ddot{\Phi}_k+\Gamma \Dot{\Phi}_k+\Omega^2\Phi_k=0.
\end{equation}
where $\Phi_k(t)$ represents the $k$-th mode of $\Phi(t,\mathbf{x})$, $\Gamma$ and $\Omega$ are the parameters depends on the function form of $f(T)$ and background evolution. In a pure gravitational system, $\Omega^2$ can be rewritten as
\begin{equation}
    \Omega^2=\frac{-\frac{f}{4}-T^2f_{TT}}{f_T+2T f_{TT}} \geq 0.
\end{equation}
For instance, a model with a negative $\Omega^2$ is obviously unstable against gravitational perturbations. To facilitate analysis, it is instructive to introduce the dimensionless perturbation variable $\omega_d^2 = -\Omega^2/T$, which retains the sign of $\Omega^2$. Fig.~\ref{fig:omega2} illustrates the values of $\omega_d^2$ obtained for our quintom model. Since in the pure gravitational scenario, the value of $T/T_0$ must be a constant between $0$ and $1$,  it becomes feasible to identify suitable parameter values that can realize quintom behavior without inducing any instability.

\section{Methodology, datasets and results} \label{sec:data&method}

\begin{figure*}[htbp]
    \centering \includegraphics[width=\textwidth]{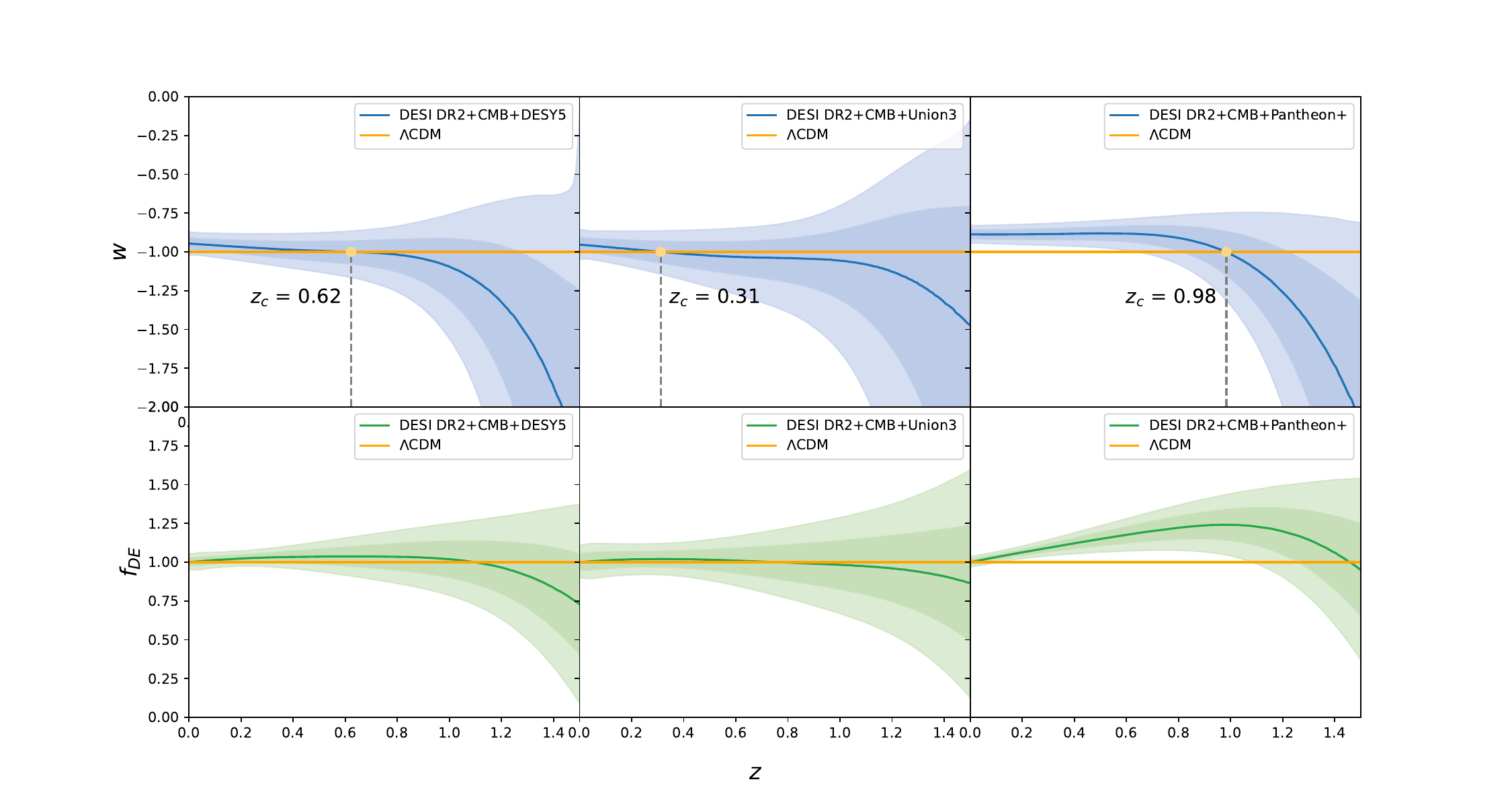}
	\caption{{\it{The mean values of the reconstructed dark energy EoS parameter $w$ and the normalized energy density $f_{de}$, along with $1\sigma$ and $2\sigma$ uncertainties, for DESI DR2+CMB+SNe datasets. For comparison, we have added the orange solid line, which corresponds to  the parameter values predicted by   $\Lambda$CDM cosmology.}}}
	\label{fig:w_gp}
\end{figure*}

\begin{figure*}[htbp]
    \centering
    \begin{minipage}{0.48\textwidth}
        \includegraphics[width=\textwidth]{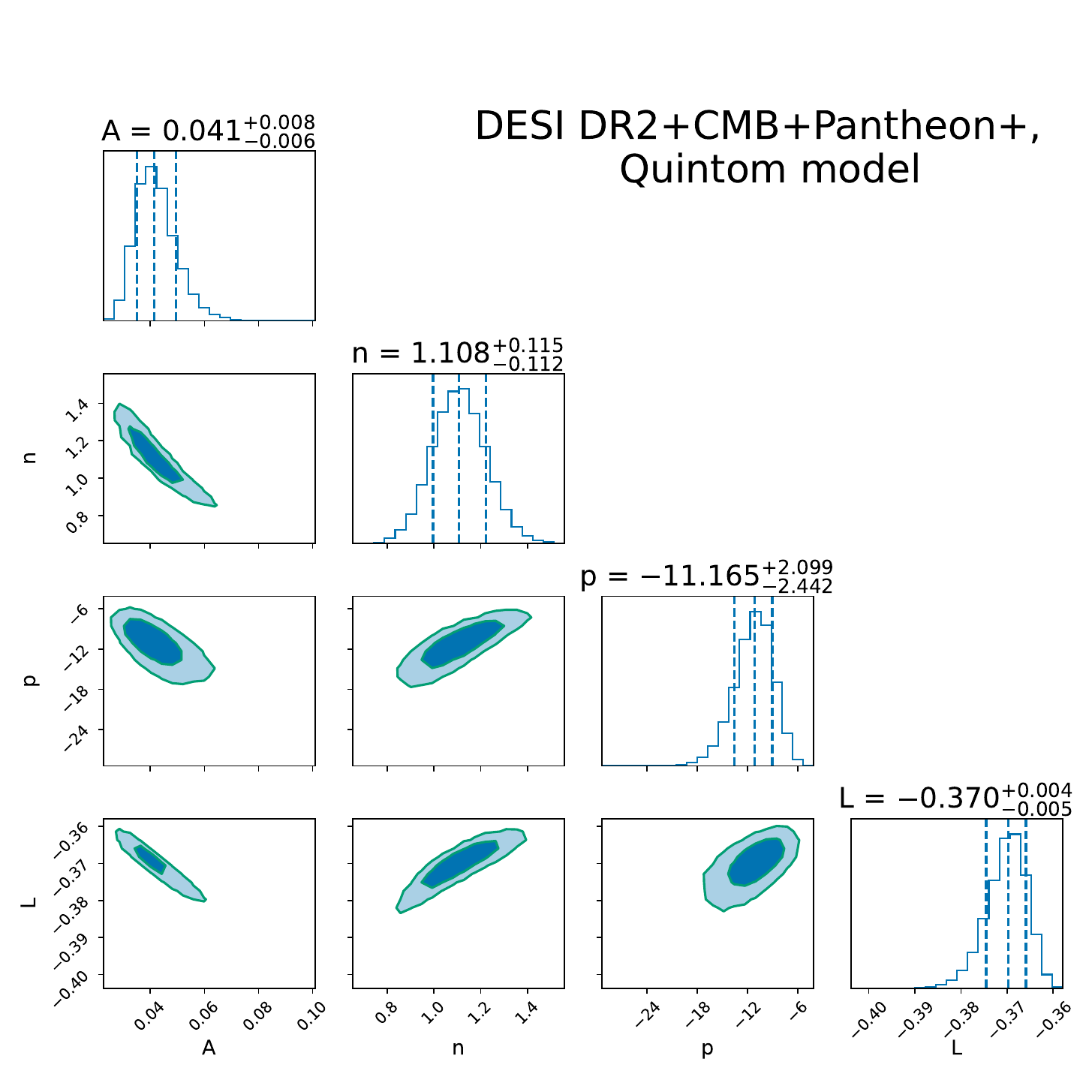}
        \label{fig:mcmc_quintom}
    \end{minipage}
    \hfill
    \begin{minipage}{0.48\textwidth}
        \includegraphics[width=\textwidth]{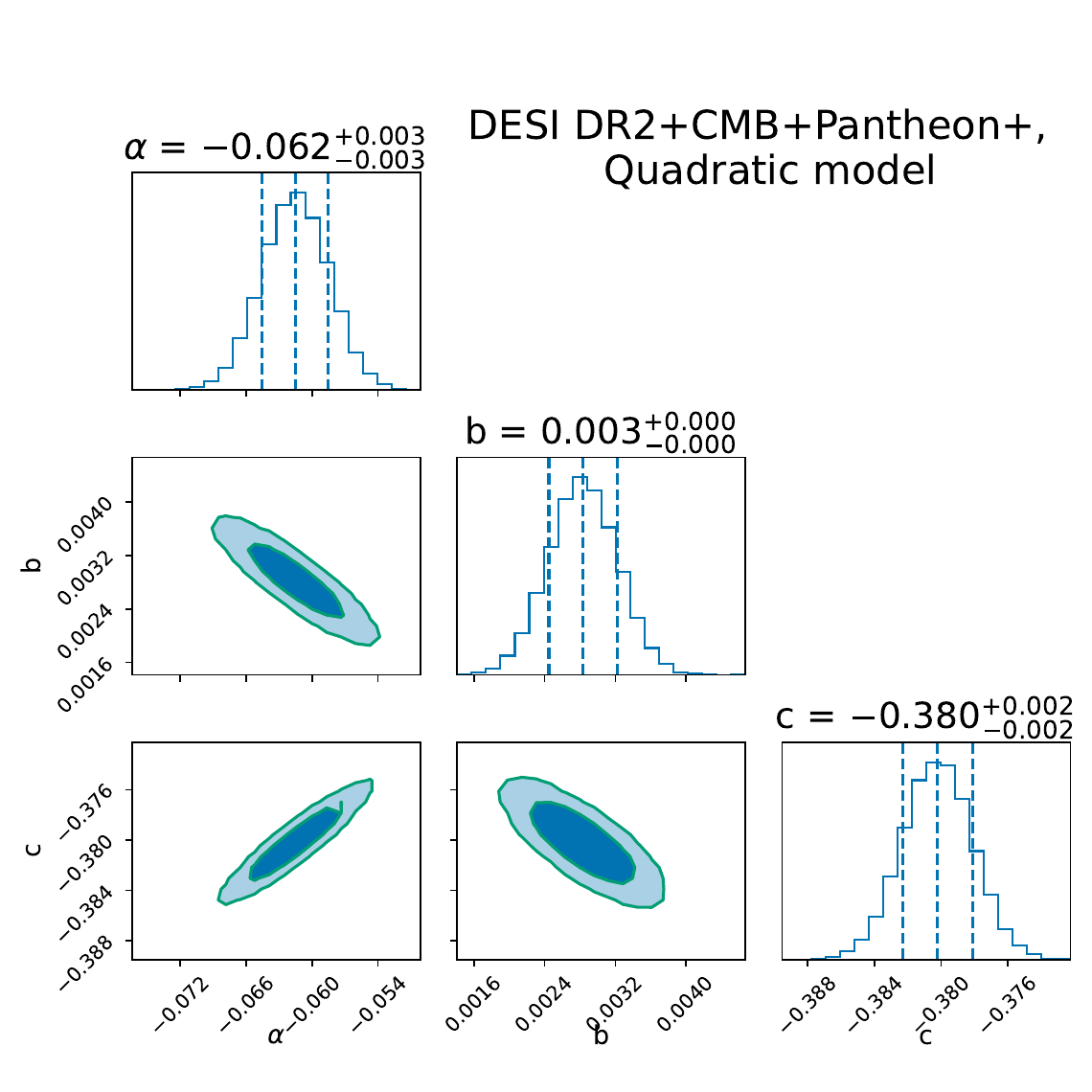}
        \label{fig:mcmc_qua}
    \end{minipage}
    \caption{\it{Marginalized 68\% and 95\% confidence level 
contours for the parameters of the  quintom model (\ref{quintommodel}) and of the quadratic model  (\ref{quadraticmodel}), from DESI DR2+CMB+Pantheon+ datasets.}}
    \label{fig:mcmc}    
\end{figure*}

In this section we proceed to the observational confrontation of the scenario at hand.
 The expansion rate can be expressed
\begin{equation}
    \frac{H(z)}{H_0}=\Big [\Omega_{m,0}(1+z)^3+\Omega_{\gamma,0}(1+z)^4+\Omega_{de,0}\frac{\rho_{de}(z)}{\rho_{de,0}}\Big ]^{1/2},
\end{equation}
with $H_0$   the current Hubble constant, and where $\Omega_{c,0}$, $\Omega_{b,0}$, $\Omega_{\gamma,0}$, $\Omega_{\nu,0}$ and $\Omega_{de,0}$ represent the energy density parameters for cold dark matter, baryons, radiation, non-relativistic massive neutrinos and dark energy at present time,  while we can  write $\Omega_m=\Omega_{c,0}+\Omega_{b,0}+\Omega_{\nu,0}$. Since radiation is not significant in the late universe,  for simplicity we set $\Omega_{\gamma,0}=0$. 

We normalize  the dark energy energy density $\rho_{de}(z)$ to its present value, leading to $f_{de}(z)\equiv \rho_{de}(z) / \rho_{de,0} $. Under the assumption that there is no interaction between the dark sectors, the normalized energy density can be expressed in terms of $w(z)$, as
\begin{equation}
    f_{de}(z)=exp\Big [ 3 \int_0^z [1+w(z')]\frac{\mathrm{d}z'}{1+z'}\Big ].
\end{equation}

Let's now discuss the datasets that we are going to use in our analysis. 
We mainly use data from three different kinds of cosmological observation. 

\textbf{SNe}--Supernovae Type Ia (SNe)   serve as standard candles due to their near-uniform peak luminosity, resulting from white dwarfs reaching the Chandrasekhar mass limit, which allow precise distance measurements via their well-calibrated light curves. In our analysis, we will utilize data from three survey compilations, namely, PantheonPlus ($0.001 < z < 2.26$) \cite{Brout:2022vxf}, Union3 ($0.05 < z < 2.26$) \cite{Rubin:2023ovl}, and DESY5 ($0.025 < z < 1.3$) \cite{Abbott:2024agi}.

\textbf{BAO}--Baryon acoustic oscillations (BAO) act as standard rulers by preserving a fixed nearly $150$ Mpc scale from primordial sound waves in the early universe, enabling geometric distance measurements through galaxy clustering patterns. We will use the latest DESI DR2 BAO results, which are divided into $7$ redshift bins and are summarized in \cite{DESI:2025zgx}. Specifically, the BAO measurements provide three different distance, namely
\begin{align}
    \frac{D_H(z)}{r_d} &= \frac{c}{H(z) r_d}, \\
    \frac{D_M(z)}{r_d} &= \int_0^z\frac{c}{H(z')r_d}\mathrm{d}z', \\
    \frac{D_V(z)}{r_d} &= \left[ z D_M^2(z) D_H(z) \right]^{1/3} / r_d,
\end{align}
where $D_H(z)$, $D_A(z)$, and $D_V(z)$ represent the Hubble distance, angular diameter distance, and volume-averaged distance, respectively. Moreover, $r_d$ denotes the sound horizon at the drag epoch, and we consider the value $r_d=147.09 \pm 0.26$ Mpc according to the Planck results \cite{Planck:2018vyg}.

\textbf{CMB}--Instead of using the full temperature and polarization measurements of the Cosmic Microwave Background (CMB), we consider the CMB measurement as an effective BAO data point at $z_* \simeq 1089$, where $z_*$ is the redshift of the photon decoupling. The angular acoustic scale $\theta_{*}$ of CMB can be expressed as
\begin{equation}
    \theta_*=r_*/D_{M}(z_*),
\end{equation}
where $r_*$ is the sound horizon at decouple time. This can be converted into a BAO observable value as
\begin{equation}
    \frac{D_M(z_*)}{r_d}=\frac{r_*}{r_d \theta_*},
\end{equation}
and
we take $100\theta_*=1.04110 \pm0.00031$ and $r_*=144.43 \pm 0.26$ Mpc \cite{Planck:2018vyg}. In what follows  we will refer to this BAO observable value simply as   CMB data.

We use the  non-parametric  Gaussian process  method  \cite{Shafieloo:2012ht,Holsclaw:2010sk,Holsclaw:2010nb,Seikel:2012uu}. 
Concerning the kernel function
we choose the squared exponential kernel covariance function, $k(x,x')=\sigma_f^2 \cdot \exp[-(x-x')^2/(2l^2)]$,   with hyperparameters $\sigma_f$ and $l$. Driven by the data, we draw a sample of $D_M$ from a multivariate Gaussian distribution with covariance matrix   the above kernel function. Then, using the relation $H(z)=c/D'_M(z)$ and background evolution equations, we are able to reconstruct $w$, $f_{de}$ and the gravitational action,   following the   procedure described in \cite{Yang:2025kgc,Yang:2024kdo,Ren:2021tfi,Cai:2019bdh}.

In Fig. \ref{fig:w_gp}  we show the reconstruction results for the dark energy EoS  $w$ and $f_{de}$. Concerning   $w$, we can see that it always crosses -1 from below, lying  on the phantom regime at high redshifts  and entering into the  quintessence regime at late times, and thus experiencing the quintom behavior. Additionally,  for $f_{de}$  we find that it tends to decrease with increasing redshift, remaining always positive.

Finally, let us investigate the $f(T)$ form itself. The quintom model \eqref{eq:quintom ft model}  can be rewritten as 
\begin{equation}
    \frac{f(T)}{T_0}=\frac{T}{T_0}-A \left (\frac{T}{T_0} \right )^{n} \times \left(1-e^{pT_0/T}\right)-2L,
        \label{quintommodel}
\end{equation}
where we have defined the dimensionless parameters $A=\alpha (-T_0)^{n-1}$ and $L=\Lambda/T_0$.  It would be helpful to introduce    the quadratic $f(T)$ model too, given by
\begin{equation}
    \frac{f(T)}{T_0}=\frac{T}{T_0}+\alpha \frac{T}{T_0}+b\frac{T^2}{T_0^2}-2c,
    \label{quadraticmodel}
\end{equation}
with $\alpha,b,c$ the model parameters. It proves convenient to 
 introduce $F(T)=f(T)-T$ in order to quantify the deviations from GR.

\begin{table}[htbp]
\centering
\begin{tabular}{ccc}
\hline
\multicolumn{3}{c} {DESI DR2+CMB+Pantheon+ datasets} \\
\hline
    Criteria& AIC & BIC\\
\hline
    Quintom model &$97.875$&$119.65$   \\ 
    Quadratic model &$97.151$ &$113.841$ \\ 
    $\Lambda$CDM model & $100.983$ & $111.87$ \\
\hline
\end{tabular}

\caption{The information criteria AIC and BIC for the quintom model (\ref{quintommodel}), the quadratic model (\ref{quadraticmodel}) {and $\Lambda$CDM model}. }
\label{table:aic}
\end{table}

{For sake of simplicity, we only focus on the data combination of DESI DR2+CMB+Pantheon+.} By using the Monte Carlo
Markov Chain method, we provide the constraints of the parameters of  both models in Fig. \ref{fig:mcmc}. Furthermore, we use the Akaike Information Criterion (AIC) and the Bayesian Information Criterion (BIC) to examine the quality of the fittings, i.e. $AIC=-2\ln{\mathcal{L}_{max}}+2p_{tot}$ and $BIC =-2\ln{\mathcal{L}_{max}}+p_{tot} \ln{N_{tot}}$, where $\ln{\mathcal{L}_{max}}$ represents the maximum likelihood of the model and
$p_{tot}$ and $N_{tot}$ represent the total number of free parameters and data points respectively \cite{Liddle:2007fy,Anagnostopoulos:2019miu}. The AIC and BIC value of  the two   models are shown in Table \ref{table:aic}. {As we can see, the quadratic model is slightly favored by AIC. But the criteria BIC prefers $\Lambda$CDM model. This is because BIC imposes a higher penalty mechanism on the number of parameters compared to AIC.} Finally, we remind that all the above results holds for $f(Q)$ gravity with the coincident gauge, under the substitution $T\rightarrow Q$.

\section{Conclusions}\label{conclusion}
The recent DESI DR2 BAO measurements provide substantial evidence for dynamical dark energy, which favors it over the standard $\Lambda$CDM paradigm with a statistical significance reaching $4.2\sigma$. Notably, the analysis reveals that the dark energy EoS parameter $w$ exhibits a redshift-dependent evolution: at  high redshifts it lies below $-1$, while at 
lower redshifts it crosses the phantom-divide, entering into the quintessence regime, which corresponds to   the quintom-B type dark energy.

In this work  we presented a   realization for the quintom-like dynamical behavior, within the framework of modified gravity theories.  Firstly, we constructed an EFT action for dark energy in a general metric-affine geometry. Then, we demonstrated how two specific modified gravity theories, namely $f(T)$ gravity and $f(Q)$ gravity, can be systematically mapped to our EFT framework. We considered a specific  $f(T)$ form that can  naturally realize both quintom-A and quintom-B dark energy dynamics through appropriate parameter choices. Using the latest observational data combined with Gaussian process reconstruction, we derived the evolution of both the dark energy EoS parameter $w(z)$ and the normalized dark energy density $f_{de}(z)$. Our reconstruction confirmed the quintom-B behavior indicated by DESI results, with $w(z)$ crossing the phantom divide ($w = -1$) from below. Furthermore, we reconstructed the  $f(T)$ form from   observations, and we     extracted the corresponding bounds on the parameters of our quintom model.

{
As the accuracy of cosmological observations gradually improves, it poses a huge challenge to the concordance cosmological model. The mounting observational evidence for dynamical dark energy that comes from the combination of BAO and SNe measurements, particularly the quintom-B scenario presents significant theoretical challenges. On the one hand,} the No-Go theorem of quintom dark energy realization in conventional scalar-field models necessitates the development of novel theoretical frameworks capable of realizing such dynamics. Due to its rich structure, modified gravity stands as one of the main candidates for the realization of dynamical dark energy, and the description of Nature. {On the other hand, we hope that future accuracy observations can help us have a better understanding about the origin of this dynamical dark energy signal and the Nature of our Universe.}

\textbf{Note added}: While our article was being finalized, another work  analyzing the results of DESI
collaboration appeared \cite{Gu:2025xie}, examining whether dark energy evolves, and showed that dark energy does indeed prefer quintom-B dynamics. The conclusions of our work are consistent with these results too, and our attention is focused on the aspects of modified gravity and can inspire more theories that could realize quintom dark energy in such a framework.

\section*{Acknowledgments}
We are grateful to Taotao Qiu, Dongdong Zhang and Xinmin Zhang for insightful comments. This work was supported in part by the National Key R\&D Program of China (2021YFC2203100,2024YFC2207500), by the National Natural Science Foundation of China (12433002, 12261131497,92476203), by CAS young interdisciplinary innovation team (JCTD-2022-20), by 111 Project (B23042), by CSC Innovation Talent Funds, by USTC Fellowship for International Cooperation, and by USTC Research Funds of the Double First-Class Initiative. 
ENS acknowledges the contribution of the LISA CosWG and the COST Actions  and 
of COST Actions CA21136 ``Addressing observational tensions in cosmology with 
systematics and fundamental physics (CosmoVerse)'',  CA21106 ``COSMIC WISPers 
in the Dark Universe: Theory, astrophysics and experiments (CosmicWISPers)'', 
and CA23130 ``Bridging high and low energies in search of quantum gravity 
(BridgeQG)''. Kavli IPMU is supported by World Premier International Research Center Initiative (WPI), MEXT, Japan.

\bibliography{main_bib}{}

\begin{thebibliography}{}
\expandafter\ifx\csname natexlab\endcsname\relax\def\natexlab#1{#1}\fi
\providecommand{\url}[1]{\href{#1}{#1}}
\providecommand{\dodoi}[1]{doi:~\href{http://doi.org/#1}{\nolinkurl{#1}}}
\providecommand{\doeprint}[1]{\href{http://ascl.net/#1}{\nolinkurl{http://ascl.net/#1}}}
\providecommand{\doarXiv}[1]{\href{https://arxiv.org/abs/#1}{\nolinkurl{https://arxiv.org/abs/#1}}}

\bibitem[{Abbott {et~al.}(2018)}]{DES:2017myr}
Abbott, T. M.~C., {et~al.} 2018, Phys. Rev. D, 98, 043526, \dodoi{10.1103/PhysRevD.98.043526}

\bibitem[{Abbott {et~al.}(2024)}]{Abbott:2024agi}
---. 2024, Astrophys. J. Lett., 973, L14, \dodoi{10.3847/2041-8213/ad6f9f}

\bibitem[{Abdul~Karim {et~al.}(2025)}]{DESI:2025zgx}
Abdul~Karim, M., {et~al.} 2025.
\newblock \doarXiv{2503.14738}

\bibitem[{Adame {et~al.}(2025{\natexlab{a}})}]{DESI:2024mwx}
Adame, A.~G., {et~al.} 2025{\natexlab{a}}, JCAP, 02, 021, \dodoi{10.1088/1475-7516/2025/02/021}

\bibitem[{Adame {et~al.}(2025{\natexlab{b}})}]{DESI:2024hhd}
---. 2025{\natexlab{b}}, JCAP, 07, 028, \dodoi{10.1088/1475-7516/2025/07/028}

\bibitem[{Ade {et~al.}(2016)}]{Planck:2015fie}
Ade, P. A.~R., {et~al.} 2016, Astron. Astrophys., 594, A13, \dodoi{10.1051/0004-6361/201525830}

\bibitem[{Aghanim {et~al.}(2020)}]{Planck:2018vyg}
Aghanim, N., {et~al.} 2020, Astron. Astrophys., 641, A6, \dodoi{10.1051/0004-6361/201833910}

\bibitem[{Anagnostopoulos {et~al.}(2019)Anagnostopoulos, Basilakos, \& Saridakis}]{Anagnostopoulos:2019miu}
Anagnostopoulos, F.~K., Basilakos, S., \& Saridakis, E.~N. 2019, Phys. Rev. D, 100, 083517, \dodoi{10.1103/PhysRevD.100.083517}

\bibitem[{Anchordoqui {et~al.}(2025)Anchordoqui, Antoniadis, \& Lust}]{Anchordoqui:2025fgz}
Anchordoqui, L.~A., Antoniadis, I., \& Lust, D. 2025, Phys. Lett. B, 868, 139632, \dodoi{10.1016/j.physletb.2025.139632}

\bibitem[{Aoki {et~al.}(2024)Aoki, Bahamonde, Gigante~Valcarcel, \& Gorji}]{Aoki:2023sum}
Aoki, K., Bahamonde, S., Gigante~Valcarcel, J., \& Gorji, M.~A. 2024, Phys. Rev. D, 110, 024017, \dodoi{10.1103/PhysRevD.110.024017}

\bibitem[{Armendariz-Picon {et~al.}(2000)Armendariz-Picon, Mukhanov, \& Steinhardt}]{Armendariz-Picon:2000nqq}
Armendariz-Picon, C., Mukhanov, V.~F., \& Steinhardt, P.~J. 2000, Phys. Rev. Lett., 85, 4438, \dodoi{10.1103/PhysRevLett.85.4438}

\bibitem[{Bahamonde {et~al.}(2023)Bahamonde, Dialektopoulos, Escamilla-Rivera, Farrugia, Gakis, Hendry, Hohmann, Levi~Said, Mifsud, \& Di~Valentino}]{Bahamonde:2021gfp}
Bahamonde, S., Dialektopoulos, K.~F., Escamilla-Rivera, C., {et~al.} 2023, Rept. Prog. Phys., 86, 026901, \dodoi{10.1088/1361-6633/ac9cef}

\bibitem[{Basilakos {et~al.}(2025)Basilakos, Paliathanasis, \& Saridakis}]{Basilakos:2025olm}
Basilakos, S., Paliathanasis, A., \& Saridakis, E.~N. 2025, Phys. Lett. B, 868, 139658, \dodoi{10.1016/j.physletb.2025.139658}

\bibitem[{Bedroya \& Vafa(2020)}]{Bedroya:2019snp}
Bedroya, A., \& Vafa, C. 2020, JHEP, 09, 123, \dodoi{10.1007/JHEP09(2020)123}

\bibitem[{Beltr\'an~Jim\'enez {et~al.}(2018)Beltr\'an~Jim\'enez, Heisenberg, \& Koivisto}]{BeltranJimenez:2017tkd}
Beltr\'an~Jim\'enez, J., Heisenberg, L., \& Koivisto, T. 2018, Phys. Rev. D, 98, 044048, \dodoi{10.1103/PhysRevD.98.044048}

\bibitem[{Beltr\'an~Jim\'enez {et~al.}(2019)Beltr\'an~Jim\'enez, Heisenberg, \& Koivisto}]{BeltranJimenez:2019esp}
Beltr\'an~Jim\'enez, J., Heisenberg, L., \& Koivisto, T.~S. 2019, Universe, 5, 173, \dodoi{10.3390/universe5070173}

\bibitem[{Beltr\'an~Jim\'enez {et~al.}(2020)Beltr\'an~Jim\'enez, Heisenberg, Koivisto, \& Pekar}]{BeltranJimenez:2019tme}
Beltr\'an~Jim\'enez, J., Heisenberg, L., Koivisto, T.~S., \& Pekar, S. 2020, Phys. Rev. D, 101, 103507, \dodoi{10.1103/PhysRevD.101.103507}

\bibitem[{Bloomfield {et~al.}(2013)Bloomfield, Flanagan, Park, \& Watson}]{Bloomfield:2012ff}
Bloomfield, J.~K., Flanagan, E.~E., Park, M., \& Watson, S. 2013, JCAP, 08, 010, \dodoi{10.1088/1475-7516/2013/08/010}

\bibitem[{Brandenberger(2025)}]{Brandenberger:2025hof}
Brandenberger, R. 2025.
\newblock \doarXiv{2503.17659}

\bibitem[{Brout {et~al.}(2022)}]{Brout:2022vxf}
Brout, D., {et~al.} 2022, Astrophys. J., 938, 110, \dodoi{10.3847/1538-4357/ac8e04}

\bibitem[{Cai {et~al.}(2016)Cai, Capozziello, De~Laurentis, \& Saridakis}]{Cai:2015emx}
Cai, Y.-F., Capozziello, S., De~Laurentis, M., \& Saridakis, E.~N. 2016, Rept. Prog. Phys., 79, 106901, \dodoi{10.1088/0034-4885/79/10/106901}

\bibitem[{Cai {et~al.}(2020)Cai, Khurshudyan, \& Saridakis}]{Cai:2019bdh}
Cai, Y.-F., Khurshudyan, M., \& Saridakis, E.~N. 2020, Astrophys. J., 888, 62, \dodoi{10.3847/1538-4357/ab5a7f}

\bibitem[{Cai {et~al.}(2010)Cai, Saridakis, Setare, \& Xia}]{Cai:2009zp}
Cai, Y.-F., Saridakis, E.~N., Setare, M.~R., \& Xia, J.-Q. 2010, Phys. Rept., 493, 1, \dodoi{10.1016/j.physrep.2010.04.001}

\bibitem[{Calderon {et~al.}(2024)}]{DESI:2024aqx}
Calderon, R., {et~al.} 2024, JCAP, 10, 048, \dodoi{10.1088/1475-7516/2024/10/048}

\bibitem[{Caldwell(2002)}]{Caldwell:1999ew}
Caldwell, R.~R. 2002, Phys. Lett. B, 545, 23, \dodoi{10.1016/S0370-2693(02)02589-3}

\bibitem[{Capozziello(2002)}]{Capozziello:2002rd}
Capozziello, S. 2002, Int. J. Mod. Phys. D, 11, 483, \dodoi{10.1142/S0218271802002025}

\bibitem[{Chen {et~al.}(2011)Chen, Dent, Dutta, \& Saridakis}]{Chen:2010va}
Chen, S.-H., Dent, J.~B., Dutta, S., \& Saridakis, E.~N. 2011, Phys. Rev. D, 83, 023508, \dodoi{10.1103/PhysRevD.83.023508}

\bibitem[{Cheung {et~al.}(2008)Cheung, Creminelli, Fitzpatrick, Kaplan, \& Senatore}]{Cheung:2007st}
Cheung, C., Creminelli, P., Fitzpatrick, A.~L., Kaplan, J., \& Senatore, L. 2008, JHEP, 03, 014, \dodoi{10.1088/1126-6708/2008/03/014}

\bibitem[{Chiba {et~al.}(2000)Chiba, Okabe, \& Yamaguchi}]{Chiba:1999ka}
Chiba, T., Okabe, T., \& Yamaguchi, M. 2000, Phys. Rev. D, 62, 023511, \dodoi{10.1103/PhysRevD.62.023511}

\bibitem[{Chimento {et~al.}(2009)Chimento, Forte, Lazkoz, \& Richarte}]{Chimento:2008ws}
Chimento, L.~P., Forte, M.~I., Lazkoz, R., \& Richarte, M.~G. 2009, Phys. Rev. D, 79, 043502, \dodoi{10.1103/PhysRevD.79.043502}

\bibitem[{Chudaykin \& Kunz(2024)}]{Chudaykin:2024gol}
Chudaykin, A., \& Kunz, M. 2024, Phys. Rev. D, 110, 123524, \dodoi{10.1103/PhysRevD.110.123524}

\bibitem[{Colg\'ain {et~al.}(2021)Colg\'ain, Sheikh-Jabbari, \& Yin}]{Colgain:2021pmf}
Colg\'ain, E.~O., Sheikh-Jabbari, M.~M., \& Yin, L. 2021, Phys. Rev. D, 104, 023510, \dodoi{10.1103/PhysRevD.104.023510}

\bibitem[{Cort\^es \& Liddle(2024)}]{Cortes:2024lgw}
Cort\^es, M., \& Liddle, A.~R. 2024, JCAP, 12, 007, \dodoi{10.1088/1475-7516/2024/12/007}

\bibitem[{Creminelli {et~al.}(2009)Creminelli, D'Amico, Norena, \& Vernizzi}]{Creminelli:2008wc}
Creminelli, P., D'Amico, G., Norena, J., \& Vernizzi, F. 2009, JCAP, 02, 018, \dodoi{10.1088/1475-7516/2009/02/018}

\bibitem[{Deffayet {et~al.}(2010)Deffayet, Pujolas, Sawicki, \& Vikman}]{Deffayet:2010qz}
Deffayet, C., Pujolas, O., Sawicki, I., \& Vikman, A. 2010, JCAP, 10, 026, \dodoi{10.1088/1475-7516/2010/10/026}

\bibitem[{Dinda(2024)}]{Dinda:2024kjf}
Dinda, B.~R. 2024, JCAP, 09, 062, \dodoi{10.1088/1475-7516/2024/09/062}

\bibitem[{Escamilla-Rivera \& Sandoval-Orozco(2024)}]{Escamilla-Rivera:2024sae}
Escamilla-Rivera, C., \& Sandoval-Orozco, R. 2024, JHEAp, 42, 217, \dodoi{10.1016/j.jheap.2024.05.005}

\bibitem[{Feng {et~al.}(2005)Feng, Wang, \& Zhang}]{Feng:2004ad}
Feng, B., Wang, X.-L., \& Zhang, X.-M. 2005, Phys. Lett. B, 607, 35, \dodoi{10.1016/j.physletb.2004.12.071}

\bibitem[{Giar{\`e}(2025)}]{Giare:2024oil}
Giar{\`e}, W. 2025, Phys. Rev. D, 112, 023508, \dodoi{10.1103/ss37-cxhn}

\bibitem[{Giar\`e {et~al.}(2024)Giar\`e, Najafi, Pan, Di~Valentino, \& Firouzjaee}]{Giare:2024gpk}
Giar\`e, W., Najafi, M., Pan, S., Di~Valentino, E., \& Firouzjaee, J.~T. 2024, JCAP, 10, 035, \dodoi{10.1088/1475-7516/2024/10/035}

\bibitem[{Gleyzes {et~al.}(2013)Gleyzes, Langlois, Piazza, \& Vernizzi}]{Gleyzes:2013ooa}
Gleyzes, J., Langlois, D., Piazza, F., \& Vernizzi, F. 2013, JCAP, 08, 025, \dodoi{10.1088/1475-7516/2013/08/025}

\bibitem[{Gronwald(1997)}]{Gronwald:1997bx}
Gronwald, F. 1997, Int. J. Mod. Phys. D, 6, 263, \dodoi{10.1142/S0218271897000157}

\bibitem[{Gu {et~al.}(2025)}]{Gu:2025xie}
Gu, G., {et~al.} 2025.
\newblock \doarXiv{2504.06118}

\bibitem[{Gubitosi {et~al.}(2013)Gubitosi, Piazza, \& Vernizzi}]{Gubitosi:2012hu}
Gubitosi, G., Piazza, F., \& Vernizzi, F. 2013, JCAP, 02, 032, \dodoi{10.1088/1475-7516/2013/02/032}

\bibitem[{Hehl {et~al.}(1995)Hehl, McCrea, Mielke, \& Ne'eman}]{Hehl:1994ue}
Hehl, F.~W., McCrea, J.~D., Mielke, E.~W., \& Ne'eman, Y. 1995, Phys. Rept., 258, 1, \dodoi{10.1016/0370-1573(94)00111-F}

\bibitem[{Heisenberg(2024)}]{Heisenberg:2023lru}
Heisenberg, L. 2024, Phys. Rept., 1066, 1, \dodoi{10.1016/j.physrep.2024.02.001}

\bibitem[{Heymans {et~al.}(2012)}]{Heymans:2012gg}
Heymans, C., {et~al.} 2012, Mon. Not. Roy. Astron. Soc., 427, 146, \dodoi{10.1111/j.1365-2966.2012.21952.x}

\bibitem[{Hohmann(2020)}]{Hohmann:2019fvf}
Hohmann, M. 2020, Symmetry, 12, 453, \dodoi{10.3390/sym12030453}

\bibitem[{Holsclaw {et~al.}(2010{\natexlab{a}})Holsclaw, Alam, Sanso, Lee, Heitmann, Habib, \& Higdon}]{Holsclaw:2010sk}
Holsclaw, T., Alam, U., Sanso, B., {et~al.} 2010{\natexlab{a}}, Phys. Rev. Lett., 105, 241302, \dodoi{10.1103/PhysRevLett.105.241302}

\bibitem[{Holsclaw {et~al.}(2010{\natexlab{b}})Holsclaw, Alam, Sanso, Lee, Heitmann, Habib, \& Higdon}]{Holsclaw:2010nb}
---. 2010{\natexlab{b}}, Phys. Rev. D, 82, 103502, \dodoi{10.1103/PhysRevD.82.103502}

\bibitem[{Horndeski(1974)}]{Horndeski:1974wa}
Horndeski, G.~W. 1974, Int. J. Theor. Phys., 10, 363, \dodoi{10.1007/BF01807638}

\bibitem[{Huang {et~al.}(2025)Huang, Cai, \& Wang}]{Huang:2025som}
Huang, L., Cai, R.-G., \& Wang, S.-J. 2025, Sci. China Phys. Mech. Astron., 68, 100413, \dodoi{10.1007/s11433-025-2754-5}

\bibitem[{Jiang {et~al.}(2024)Jiang, Pedrotti, da~Costa, \& Vagnozzi}]{Jiang:2024xnu}
Jiang, J.-Q., Pedrotti, D., da~Costa, S.~S., \& Vagnozzi, S. 2024, Phys. Rev. D, 110, 123519, \dodoi{10.1103/PhysRevD.110.123519}

\bibitem[{Jim\'enez~Cano(2021)}]{JimenezCano:2021rlu}
Jim\'enez~Cano, A. 2021, PhD thesis, Granada U., Theor. Phys. Astrophys.
\newblock \doarXiv{2201.12847}

\bibitem[{Krssak {et~al.}(2019)Krssak, van~den Hoogen, Pereira, B\"ohmer, \& Coley}]{Krssak:2018ywd}
Krssak, M., van~den Hoogen, R., Pereira, J., B\"ohmer, C., \& Coley, A. 2019, Class. Quant. Grav., 36, 183001, \dodoi{10.1088/1361-6382/ab2e1f}

\bibitem[{Langlois {et~al.}(2017)Langlois, Mancarella, Noui, \& Vernizzi}]{Langlois:2017mxy}
Langlois, D., Mancarella, M., Noui, K., \& Vernizzi, F. 2017, JCAP, 05, 033, \dodoi{10.1088/1475-7516/2017/05/033}

\bibitem[{Langlois {et~al.}(2019)Langlois, Mancarella, Noui, \& Vernizzi}]{Langlois:2018jdg}
---. 2019, JCAP, 02, 036, \dodoi{10.1088/1475-7516/2019/02/036}

\bibitem[{Li {et~al.}(2018)Li, Cai, Cai, \& Saridakis}]{Li:2018ixg}
Li, C., Cai, Y., Cai, Y.-F., \& Saridakis, E.~N. 2018, JCAP, 10, 001, \dodoi{10.1088/1475-7516/2018/10/001}

\bibitem[{Li {et~al.}(2025)Li, Wang, Zhang, Saridakis, \& Cai}]{Li:2025cxn}
Li, C., Wang, J., Zhang, D., Saridakis, E.~N., \& Cai, Y.-F. 2025.
\newblock \doarXiv{2504.07791}

\bibitem[{Liddle(2007)}]{Liddle:2007fy}
Liddle, A.~R. 2007, Mon. Not. Roy. Astron. Soc., 377, L74, \dodoi{10.1111/j.1745-3933.2007.00306.x}

\bibitem[{Linder(2010)}]{Linder:2010py}
Linder, E.~V. 2010, Phys. Rev. D, 81, 127301, \dodoi{10.1103/PhysRevD.81.127301}

\bibitem[{Liu {et~al.}(2024)Liu, Wang, \& Zhao}]{Liu:2024gfy}
Liu, G., Wang, Y., \& Zhao, W. 2024.
\newblock \doarXiv{2407.04385}

\bibitem[{Lodha {et~al.}(2025{\natexlab{a}})}]{DESI:2024kob}
Lodha, K., {et~al.} 2025{\natexlab{a}}, Phys. Rev. D, 111, 023532, \dodoi{10.1103/PhysRevD.111.023532}

\bibitem[{Lodha {et~al.}(2025{\natexlab{b}})}]{Lodha:2025qbg}
---. 2025{\natexlab{b}}.
\newblock \doarXiv{2503.14743}

\bibitem[{Mukherjee \& Sen(2024)}]{Mukherjee:2024ryz}
Mukherjee, P., \& Sen, A.~A. 2024, Phys. Rev. D, 110, 123502, \dodoi{10.1103/PhysRevD.110.123502}

\bibitem[{Nester \& Yo(1999)}]{Nester:1998mp}
Nester, J.~M., \& Yo, H.-J. 1999, Chin. J. Phys., 37, 113.
\newblock \doarXiv{gr-qc/9809049}

\bibitem[{Ormondroyd {et~al.}(2025)Ormondroyd, Handley, Hobson, \& Lasenby}]{Ormondroyd:2025iaf}
Ormondroyd, A.~N., Handley, W.~J., Hobson, M.~P., \& Lasenby, A.~N. 2025.
\newblock \doarXiv{2503.17342}

\bibitem[{Paliathanasis(2025)}]{Paliathanasis:2025dcr}
Paliathanasis, A. 2025.
\newblock \doarXiv{2503.20896}

\bibitem[{Paliathanasis {et~al.}(2024)Paliathanasis, Dimakis, \& Christodoulakis}]{Paliathanasis:2023pqp}
Paliathanasis, A., Dimakis, N., \& Christodoulakis, T. 2024, Phys. Dark Univ., 43, 101410, \dodoi{10.1016/j.dark.2023.101410}

\bibitem[{Pan \& Ye(2025)}]{Pan:2025psn}
Pan, J., \& Ye, G. 2025.
\newblock \doarXiv{2503.19898}

\bibitem[{Pan {et~al.}(2025)Pan, Paul, Saridakis, \& Yang}]{Pan:2025qwy}
Pan, S., Paul, S., Saridakis, E.~N., \& Yang, W. 2025.
\newblock \doarXiv{2504.00994}

\bibitem[{Pang {et~al.}(2025)Pang, Zhang, \& Huang}]{Pang:2025lvh}
Pang, Y.-H., Zhang, X., \& Huang, Q.-G. 2025, Sci. China Phys. Mech. Astron., 68, 280410, \dodoi{10.1007/s11433-025-2713-8}

\bibitem[{Park {et~al.}(2010)Park, Zurek, \& Watson}]{Park:2010cw}
Park, M., Zurek, K.~M., \& Watson, S. 2010, Phys. Rev. D, 81, 124008, \dodoi{10.1103/PhysRevD.81.124008}

\bibitem[{Perlmutter {et~al.}(1999)}]{SupernovaCosmologyProject:1998vns}
Perlmutter, S., {et~al.} 1999, Astrophys. J., 517, 565, \dodoi{10.1086/307221}

\bibitem[{Piazza \& Vernizzi(2013)}]{Piazza:2013coa}
Piazza, F., \& Vernizzi, F. 2013, Class. Quant. Grav., 30, 214007, \dodoi{10.1088/0264-9381/30/21/214007}

\bibitem[{Pogosian {et~al.}(2022)Pogosian, Raveri, Koyama, Martinelli, Silvestri, Zhao, Li, Peirone, \& Zucca}]{Pogosian:2021mcs}
Pogosian, L., Raveri, M., Koyama, K., {et~al.} 2022, Nature Astron., 6, 1484, \dodoi{10.1038/s41550-022-01808-7}

\bibitem[{Ratra \& Peebles(1988)}]{Ratra:1987rm}
Ratra, B., \& Peebles, P. J.~E. 1988, Phys. Rev. D, 37, 3406, \dodoi{10.1103/PhysRevD.37.3406}

\bibitem[{Ren {et~al.}(2021)Ren, Wong, Cai, \& Saridakis}]{Ren:2021tfi}
Ren, X., Wong, T. H.~T., Cai, Y.-F., \& Saridakis, E.~N. 2021, Phys. Dark Univ., 32, 100812, \dodoi{10.1016/j.dark.2021.100812}

\bibitem[{Ren {et~al.}(2022)Ren, Yan, Zhao, Cai, \& Saridakis}]{Ren:2022aeo}
Ren, X., Yan, S.-F., Zhao, Y., Cai, Y.-F., \& Saridakis, E.~N. 2022, Astrophys. J., 932, 131, \dodoi{10.3847/1538-4357/ac6ba5}

\bibitem[{Riess {et~al.}(1998)}]{SupernovaSearchTeam:1998fmf}
Riess, A.~G., {et~al.} 1998, Astron. J., 116, 1009, \dodoi{10.1086/300499}

\bibitem[{Roy~Choudhury \& Okumura(2024)}]{RoyChoudhury:2024wri}
Roy~Choudhury, S., \& Okumura, T. 2024, Astrophys. J. Lett., 976, L11, \dodoi{10.3847/2041-8213/ad8c26}

\bibitem[{Rubin {et~al.}(2023)}]{Rubin:2023ovl}
Rubin, D., {et~al.} 2023.
\newblock \doarXiv{2311.12098}

\bibitem[{Saridakis {et~al.}(2021)}]{CANTATA:2021asi}
Saridakis, E.~N., {et~al.} 2021, {Modified Gravity and Cosmology. An Update by the CANTATA Network}, ed. E.~N. Saridakis, R.~Lazkoz, V.~Salzano, P.~Vargas~Moniz, S.~Capozziello, J.~Beltr\'an~Jim\'enez, M.~De~Laurentis, \& G.~J. Olmo (Springer), \dodoi{10.1007/978-3-030-83715-0}

\bibitem[{Seikel {et~al.}(2012)Seikel, Clarkson, \& Smith}]{Seikel:2012uu}
Seikel, M., Clarkson, C., \& Smith, M. 2012, JCAP, 06, 036, \dodoi{10.1088/1475-7516/2012/06/036}

\bibitem[{Shafieloo {et~al.}(2012)Shafieloo, Kim, \& Linder}]{Shafieloo:2012ht}
Shafieloo, A., Kim, A.~G., \& Linder, E.~V. 2012, Phys. Rev. D, 85, 123530, \dodoi{10.1103/PhysRevD.85.123530}

\bibitem[{Starobinsky(1980)}]{Starobinsky:1980te}
Starobinsky, A.~A. 1980, Phys. Lett. B, 91, 99, \dodoi{10.1016/0370-2693(80)90670-X}

\bibitem[{Tsujikawa(2015)}]{Tsujikawa:2014mba}
Tsujikawa, S. 2015, Lect. Notes Phys., 892, 97, \dodoi{10.1007/978-3-319-10070-8_4}

\bibitem[{Vikman(2005)}]{Vikman:2004dc}
Vikman, A. 2005, Phys. Rev. D, 71, 023515, \dodoi{10.1103/PhysRevD.71.023515}

\bibitem[{Wang \& Piao(2024)}]{Wang:2024dka}
Wang, H., \& Piao, Y.-S. 2024.
\newblock \doarXiv{2404.18579}

\bibitem[{Wetterich(1988)}]{Wetterich:1987fm}
Wetterich, C. 1988, Nucl. Phys. B, 302, 668, \dodoi{10.1016/0550-3213(88)90193-9}

\bibitem[{Wu {et~al.}(2024)Wu, Ren, Yang, Hu, \& Saridakis}]{Wu:2024vcr}
Wu, C., Ren, X., Yang, Y., Hu, Y.-M., \& Saridakis, E.~N. 2024.
\newblock \doarXiv{2412.01104}

\bibitem[{Yan {et~al.}(2020)Yan, Zhang, Chen, Zhang, Cai, \& Saridakis}]{Yan:2019gbw}
Yan, S.-F., Zhang, P., Chen, J.-W., {et~al.} 2020, Phys. Rev. D, 101, 121301, \dodoi{10.1103/PhysRevD.101.121301}

\bibitem[{Yang {et~al.}(2024{\natexlab{a}})Yang, Ren, Wang, Cai, \& Saridakis}]{Yang:2024tkw}
Yang, Y., Ren, X., Wang, B., Cai, Y.-F., \& Saridakis, E.~N. 2024{\natexlab{a}}, Mon. Not. Roy. Astron. Soc., 533, 2232, \dodoi{10.1093/mnras/stae1905}

\bibitem[{Yang {et~al.}(2024{\natexlab{b}})Yang, Ren, Wang, Lu, Zhang, Cai, \& Saridakis}]{Yang:2024kdo}
Yang, Y., Ren, X., Wang, Q., {et~al.} 2024{\natexlab{b}}, Sci. Bull., 69, 2698, \dodoi{10.1016/j.scib.2024.07.029}

\bibitem[{Yang {et~al.}(2025)Yang, Wang, Li, Yuan, Ren, Saridakis, \& Cai}]{Yang:2025kgc}
Yang, Y., Wang, Q., Li, C., {et~al.} 2025.
\newblock \doarXiv{2501.18336}

\bibitem[{Yin(2024)}]{Yin:2024hba}
Yin, W. 2024, JHEP, 05, 327, \dodoi{10.1007/JHEP05(2024)327}

\bibitem[{Zhao {et~al.}(2012)Zhao, Crittenden, Pogosian, \& Zhang}]{Zhao:2012aw}
Zhao, G.-B., Crittenden, R.~G., Pogosian, L., \& Zhang, X. 2012, Phys. Rev. Lett., 109, 171301, \dodoi{10.1103/PhysRevLett.109.171301}

\bibitem[{Zhao {et~al.}(2017)}]{Zhao:2017cud}
Zhao, G.-B., {et~al.} 2017, Nature Astron., 1, 627, \dodoi{10.1038/s41550-017-0216-z}

\end{thebibliography}
\bibliographystyle{aasjournal}

\end{document}